\documentclass[fleqn,usenatbib]{mnras}

\usepackage{newtxtext,newtxmath}

\usepackage[T1]{fontenc}

\DeclareRobustCommand{\VAN}[3]{#2}
\let\VANthebibliography\thebibliography
\def\thebibliography{\DeclareRobustCommand{\VAN}[3]{##3}\VANthebibliography}

\usepackage{graphicx}	
\usepackage{amsmath}	
\usepackage{tikz}
\usepackage{float}
\usepackage{subcaption}
\usepackage{geometry}

\title[Representation learning for FRB spectra]{Representation learning for fast radio burst dynamic spectra}

\author[D. Kuiper et al.]{
Dirk Kuiper,$^{1}$\thanks{E-mail: d.kuiper@uva.nl}
Gabriella Contardo,$^{2,7}$
Daniela Huppenkothen,$^{1,3}$
Jason W. T. Hessels$^{1,4,5,6}$
\\
$^{1}$Anton Pannekoek Institute for Astronomy, University of Amsterdam, Science Park 904, NL-1098 XH Amsterdam, The Netherlands\\
$^{2}$Center for Astrophysics and Cosmology, University of Nova Gorica, Vipavska 11c, SI-5270 Ajdovščina, Slovenia\\
$^{3}$ SRON Netherlands Institute for Space Research, Niels Bohrlaan 4, 2333CA Leiden, The Netherlands\\
$^{4}$ASTRON, Netherlands Institute for Radio Astronomy, Oude Hoogeveensedijk 4, 7991 PD Dwingeloo, The Netherlands\\
$^{5}$Trottier Space Institute, McGill University, 3550 rue University, Montréal, QC H3A 2A7, Canada\\
$^{6}$Department of Physics, McGill University, 3600 rue University, Montréal, QC H3A 2T8, Canada\\
$^{7}$Theoretical and Scientific Data Science, Scuola Internazionale Superiore di Studi Avanzati (SISSA), Via Bonomea, 265, 34136 Trieste TS, Italy\\}

\date{Accepted XXX. Received YYY; in original form ZZZ}

\pubyear{2024}

\begin{document}
\label{firstpage}
\pagerange{\pageref{firstpage}--\pageref{lastpage}}
\maketitle

\begin{abstract}
Fast radio bursts (FRBs) are millisecond-duration radio transients of extragalactic origin, with diverse time-frequency patterns and emission properties that require explanation. With one possible exception, FRBs are detected only in the radio, analyzing their dynamic spectra is therefore crucial to disentangling the physical processes governing their generation and propagation. Furthermore, comparing FRB morphologies provides insights into possible differences among their progenitors and environments. This study applies unsupervised learning and deep learning techniques to investigate FRB dynamic spectra, focusing on two approaches: Principal Component Analysis (PCA) and a Convolutional Autoencoder (CAE) enhanced by an Information-Ordered Bottleneck (IOB) layer. PCA served as a computationally efficient baseline, capturing broad trends, identifying outliers, and providing valuable insights into large datasets. However, its linear nature limited its ability to reconstruct complex FRB structures. In contrast, the IOB-augmented CAE excelled at capturing intricate features, with high reconstruction accuracy and effective denoising at modest signal-to-noise ratios. The IOB layer’s ability to prioritize relevant features enabled efficient data compression, preserving key morphological characteristics with minimal latent variables. When applied to real FRBs from CHIME, the IOB-CAE generalized effectively, revealing a latent space that highlighted the continuum of FRB morphologies and the potential for distinguishing intrinsic differences between burst types. This framework demonstrates that while FRBs may not naturally cluster into discrete groups, advanced representation learning techniques can uncover meaningful structures, offering new insights into the diversity and origins of these bursts.
\end{abstract}

\begin{keywords}
fast radio bursts -- machine learning -- simulations -- data analysis
\end{keywords}



\section{Introduction}

Fast radio bursts (FRBs) are enigmatic, extragalactic radio transients of unknown physical origin \citep{FRBreview, Cordes_2019, review2}. They are characterized by their short but wide range of durations -- from microsecond-scale bursts \citep{snelders_2023} to the longest-observed burst lasting up to 3 seconds, with strictly periodic sub-bursts \citep{Chime}. They also exhibit high dispersion measures (DMs), exceeding the maximum expected from our Galaxy and confirming their extragalactic origins \citep{Thornton2013APO,Connor2018OnDR, Xu2015ExtragalacticDM}. Since their discovery in 2007 \citep{lorimer2007}, FRBs have generated widespread interest due to their potential to probe astrophysical and cosmological phenomena \citep[e.g.,][]{macqart2020, Zhou2014, Bannister2019ASF, Ravi2016TheMF, shin2024}. 

Key open questions concern the progenitors of FRBs, i.e. the astrophysical objects or systems that produce these bursts, as well as their radio emission mechanisms, i.e. what physics produces the bursts themselves. The distinction between repeaters and non-repeaters may indicate at least two different burst types, even if they arise from the same progenitor \citep{kirsten2023connectingrepeatingnonrepeatingfast}. Repeaters, such as FRB~20121102A \citep{spitler2016}, exhibit recurring bursts with sometimes complex morphologies and frequency drifts \citep{Hessels_2019}. They are generally wider in time but narrower in bandwidth, while non-repeaters are characterized by isolated, single events, that are narrower in time, but wider in bandwidth, suggesting potentially different physical conditions or environments \citep{Pleunis_2021}.

While multi-wavelength counterparts to extragalactic FRBs remain elusive \citep{Chen2020TheMC, Pearlman_2024}, efforts to unravel the progenitors and emission mechanisms of FRBs can take two complementary approaches: investigating their host galaxies and local environments, or analyzing the properties of the radio bursts themselves.\footnote{A notable exception is the Galactic magnetar SGR 1935+2154, which produced a bright burst that is FRB-like (though less luminous) and was accompanied by an X-ray burst \citep{2020,Bochenek2020,Mereghetti2020}}. This paperfocuses on the second approach.

At the highest level of detail, FRB data are captured as voltage data—Nyquist-sampled streams for both polarization channels \citep[e.g.,][]{Price_2019, nimmo22}. These data are often converted into dynamic spectra, trading time and frequency resolution to balance the focus between fine-scale features in frequency or time. Dynamic spectra can represent the total intensity (Stokes I) as well as polarization properties (Stokes Q, U, V), although polarimetric information is not always available. Further processing often distills dynamic spectra into timeseries data or fitted parameters, potentially obscuring some of the detailed information available in the raw data. Dispersion correction further complicates matters, as it involves fitting for the dispersion measure (DM) to account for a quadratic time-frequency delay. This process can introduce ambiguities because there is a degeneracy between the DM and intrinsic changes in the burst morphology with frequency, as well as time-frequency drifts that apparently arise from the emission process itself rather than propagation through the intervening magneto-ionized media \citep{Hessels_2019}.

Propagation effects also influence the observed morphology of FRBs, even after dedispersion is applied. Scintillation introduces fine frequency structure; scattering causes asymmetric pulse broadening; and Faraday rotation modulates polarization, together encoding information about the burst's journey through the intervening ionized and magnetised plasma \citep{Nimmo2021HighlyPM, Hessels_2019, pandhi2024polarizationproperties128nonrepeating, shin2024}.  These effects can obscure intrinsic burst features while introducing new features that are not related to the emission process itself, complicating the interpretation of the burst's underlying properties.

Over 800 distinct FRB sources have been detected to date\footnote{\url{https://www.wis-tns.org/} \& \url{https://blinkverse.zero2x.org}}, and high-time-resolution studies of some of these bursts have revealed a wide variety of FRB morphologies. Morphological diversity encompasses features such as burst duration, polarization, spectral bandwidth, and temporal structures like micro-bursts, drifting patterns in time-frequency, or periodic sub-bursts. These characteristics provide valuable clues about FRB progenitors and environments, pointing to a range of emission mechanisms and propagation effects \citep{microshots, Nimmo2021HighlyPM, Majid2021ABF, Hessels_2019}. Repeaters, in particular, often display more complex, multi-component bursts compared to one-off events \citep{Pleunis_2021}.

For instance, FRB~20121102A exhibits complex time–frequency behavior, including sub-bursts with frequency-dependent drift rates and varying bandwidths. This so-called `sad trombone' effect suggests radius-to-frequency mapping, where the burst emission shifts to lower radio frequencies as the emission region propagates further from the central engine, encountering regions of lower plasma density and magnetic field strength \citep{Hessels_2019}. Similarly, on sub-burst timescales, quasi-periodic sub-structure has been reported for several FRBs, and provides a potential connection to phenomena seen in pulsar magnetospheres \citep{kramer2024}.

FRB~170827 provides another example with its $\sim$30\,$\mu$s microstructure, which constrains the emission region to be less than $\sim$10\,km in size, illustrating the potential for FRBs to probe extreme astrophysical environments \citep{farah}. The temporal and spectral modulation observed in this burst is consistent with a combination of intrinsic source properties and propagation effects, such as scattering. Furthermore, ASKAP studies have revealed diverse polarization and scattering properties among localized FRBs, highlighting the role of the circumburst medium in shaping observed properties. For example, FRB~20190711A exhibited three distinct sub-bursts with consistent rotation measures but varying polarization properties, characteristic of repeating sources \citep{Day_2020}.

Additionally, FRB~20200120E, localised to a globular cluster in the M81 galactic system, adds further morphological diversity with sub-100-nanosecond structures separated by $2-3$\,$\mu$s, suggesting a progenitor distinct from those associated with core-collapse supernovae \citep{Majid2021ABF, Kirsten2022, nimmo22}.

A recent CHIME/FRB study of twelve bright, complex FRBs highlighted a variety of morphological traits, including microstructure features as narrow as $\sim$7 $\mu$s and drifting patterns deviating from the linear drift typically observed in repeaters. These observations, combined with varied polarization properties and Faraday rotation measures, support models involving relativistic shocks, magnetospheric activity, or propagation through ionized plasma structures \citep{faber2023morphologiesbrightcomplexfast}.

Despite these advances, the majority of FRBs remain poorly categorized due to the lack of high-quality, raw voltage data. Instruments like CHIME have contributed significantly to the known FRB population, discovering over 500 new sources in its first year of operation alone and now accounting for the majority of detected FRBs \citep{chimecat1}. However, most CHIME/FRB detections rely on intensity data with a time resolution of only $\sim$1\,ms, whereas the voltage data, available for only a subset of bursts, provide a much finer time resolution of $\sim$2.5 $\mu$s—approximately 40 times better \citep{Michilli_2021}. This limitation in available high-resolution data likely leads to an underrepresentation of the full diversity of FRB morphologies in the current sample. Expanding methodologies to analyze dynamic spectra more comprehensively and leveraging high-time-resolution data where possible are therefore essential for uncovering and understanding the wide range of FRB morphologies, which is a central goal of this work.

Furthermore, the future of FRB research is marked by the expectation of increasingly large datasets. Current estimates suggest that the all-sky FRB rate is on the order of thousands per day above 1 Jy ms \citep{Champion2015FiveNF, Agarwal2020InitialRF,bhandari2018, FRBreview}, and with the continued development of detection technologies and next-generation telescopes, the rate of detection is expected to soar. Instruments such as the Square Kilometre Array (SKA) \citep{macquart2015fasttransientscosmologicaldistances}, the Deep Synoptic Array (DSA-2000) \citep{DSA}, and the Bustling Universe Radio Survey Telescope in Taiwan (BURSTT) \citep{Lin_2022}, are poised to contribute significantly once fully operational. Additionally, CHORD \citep{chord}, building on the success of CHIME, which has already made significant contributions to FRB detection, will enhance discovery rates with improved sensitivity and sky coverage. Together with the upcoming LOFAR2.0 upgrade, which will expand capabilities at lower frequencies, these instrruments will operate across a wide range of frequency bands from MHz up to GHz frequencies and open new windows to understand FRB emission mechanisms \citep{Pearlman2020MultiwavelengthRO, Ravi2018ExplainingTS, Collaboration2019ObservationsOF}. Moreover, the five-hundred-meter aperture spherical telescope (FAST; \citealt{NAN_2011}) is already detecting thousands of bursts from hyperactive repeaters \citep[e.g.,][]{Zhang2022FASTOO}. This represents both an opportunity and a challenge. While larger datasets will enable more detailed population studies, the sheer volume of data will make manual analysis impractical. Machine learning will play an increasingly important role in addressing these challenges by enabling efficient processing and classification of the growing diversity in FRB observations.

Unsupervised machine learning has already been applied in FRB research, primarily to address specific classification tasks such as distinguishing repeaters from non-repeaters. These approaches typically rely on inferred parameters like pulse widths, spectral indices, and dispersion measures \citep{Sharma_2024, sun2024exploringkeyfeaturesrepeating}, which, while possibly effective for such tasks, fail to capture the detailed morphologies of FRBs. Key features, such as sub-burst structure and frequency drift, which are critical for understanding the physical mechanisms behind FRBs, are often absent from these simplified representations.

Techniques like Uniform Manifold Approximation and Projection (UMAP) \citep{mcinnes2020umapuniformmanifoldapproximation}, t-Distributed Stochastic Neighbor Embedding (t-SNE) \citep{Maaten2008VisualizingDU}, and Random Forest classifiers \citep{Svetnik2003RandomFA} have been used to classify FRBs based on these inferred characteristics \citep{Chen2021UncloakingHR, Luo2022MachineLC, ZhuGe2022MachineLC}. UMAP and t-SNE are non-linear dimensionality reduction techniques that visualize high-dimensional data by preserving relationships between data points. Random Forest classifiers are ensemble models that combine multiple decision trees to improve classification accuracy. However, these methods focus on inferred simplified parameters and fail to capture the rich temporal and spectral details of raw dynamic spectra.

In this paper, our goal is to develop an unsupervised approach for grouping FRBs according to their time-frequency morphology, in order to systematically study their diverse structures without relying on manual classification. By analyzing FRB morphologies in this way, we aim to rediscover known classes and potentially identify new ones, shedding light on the physical mechanisms driving these bursts. Unlike the previous approaches that depend on inferred parameters, we focus on the Stokes~I dynamic spectra, which contain rich temporal and spectral information and are inherently high-dimensional. In the context of this work, the dimensionality of the dynamic spectra is defined as the number of frequency channels multiplied by the number of time samples used to represent the dynamic spectrum. Each burst's dynamic spectrum captures its variation in intensity as a function of both time and frequency, leading to datasets with hundreds of thousands of dimensions. 

To address this complexity, we seek methods that can project these dynamic spectra into a lower-dimensional space, where the most relevant features of the data are retained, and redundant or less informative variations are removed. This potentially makes it easier to identify meaningful patterns and relationships in the data.

To achieve this, we explore two distinct methods for analyzing FRB dynamic spectra: a linear approach using Principal Component Analysis (PCA; \citealt{Wold1987PrincipalCA}) and a more advanced, non-linear method, the Information-Ordered Bottleneck (IOB;  \citealt{ho2023informationordered}). PCA is a widely used technique for dimensionality reduction that identifies the directions in the data that capture the largest variations (the `principal components'). This allows the data to be summarized with fewer variables while retaining its most significant features. One advantage of PCA is its simplicity and interpretability; it provides a straightforward way to visualize and understand the most prominent patterns in the data, making it accessible and useful for exploratory analysis. However, its limited ability to capture non-linear relationships can be problematic when analyzing complex datasets like FRB dynamic spectra, where such relationships often provide important insights. In contrast, the IOB, a neural network-based non-linear technique, is designed to preserve critical information while achieving more efficient compression of the data. This compression is not just about reducing data size but also about organizing the data into a more manageable form where key patterns and structures are more apparent. By isolating the most relevant features and discarding noise or redundant information, the IOB enables one to uncover intricate patterns in the dynamic spectra that might otherwise be hidden in the high-dimensional raw data, making it particularly suitable for studying FRB morphologies in detail.

We apply these methods to explore two key aspects of FRB morphology analysis: i. the ability to compress high-dimensional data into a meaningful lower-dimensional representation while retaining key morphological features, and; ii. how effectively the lower dimensional representation generated by these methods group similar morphologies and highlight potential outliers. To test these aspects, we utilize simulations of synthetic FRB data to evaluate the performance of both PCA and the IOB, enabling a controlled comparison of their ability to represent known morphology types. This is needed to understand whether these methods can reliably capture the diversity of FRB structures in real data.

Additionally, we validate our findings using Stokes I spectra, generated from complex voltage data from CHIME \citep{CHIMEFRB:baseband}, leveraging its high-time-resolution capabilities to test how well the latent spaces generated by PCA and the IOB generalize to real observational data. Through this dual approach of simulations and observational data, we assess the scalability and robustness of these methods in analyzing FRB morphologies.

The structure of our paper is as follows: Section~\ref{sec:simulation_and_data} details the simulations and data used in our analysis, specifically focusing on the dynamic spectra. Section~\ref{sec: algos} describes our methodology for developing and evaluating unsupervised representations, while Sections~\ref{sec: results1} and \ref{sec: results2} present the results of our analysis and discusses the implications of our findings for the broader study of FRBs. Finally, Section~\ref{sec:conclusions_outlook} concludes by summarizing our work and presenting the potential future directions.

\section{Simulation framework and Data}
\label{sec:simulation_and_data}

\subsection{FRBakery: A Simulation Tool for Synthetic FRBs}
\label{sec:frbakery}

Understanding the capabilities and limitations of machine learning algorithms when analyzing FRB data requires controlled datasets with known properties. To this end, we developed a simulation framework, \texttt{FRBakery}\footnote{The full code for FRBakery can be found at: \url{https://github.com/SRON-API-DataMagic/Rep_Learn_FRB/tree/main/FRBakery}}, to generate synthetic FRB dynamic spectra. This tool, based on the \texttt{WILL} Python package \citep{kania_will_2023}, enables us to simulate a variety of FRB morphologies described by \citet{Pleunis_2021} (simple broadband, simple narrowband, temporally complex, downward drifting), as well as an additional category for scattered bursts.

The motivation for creating synthetic data is not solely a lack of available real observations but also the need to systematically test and understand what algorithms can (and cannot) learn from FRB dynamic spectra in the absence of ground-truth information for real observations. Furthermore, the available data on bursts is not evenly spread across categories, limiting the ability to robustly train and evaluate machine learning models. By using simulated data with precisely known properties, we can assess the ability of these models to differentiate FRB types, quantify their performance, and identify potential failure points. This is particularly important for unsupervised algorithms, where a thorough understanding of their behavior is critical to effectively interpret (and avoid overinterpreting) their results on real data \citep{rudin2021interpretablemachinelearningfundamental, Chen2023Machine}.

\begin{figure*}
    \centering
        \includegraphics[width=\textwidth]{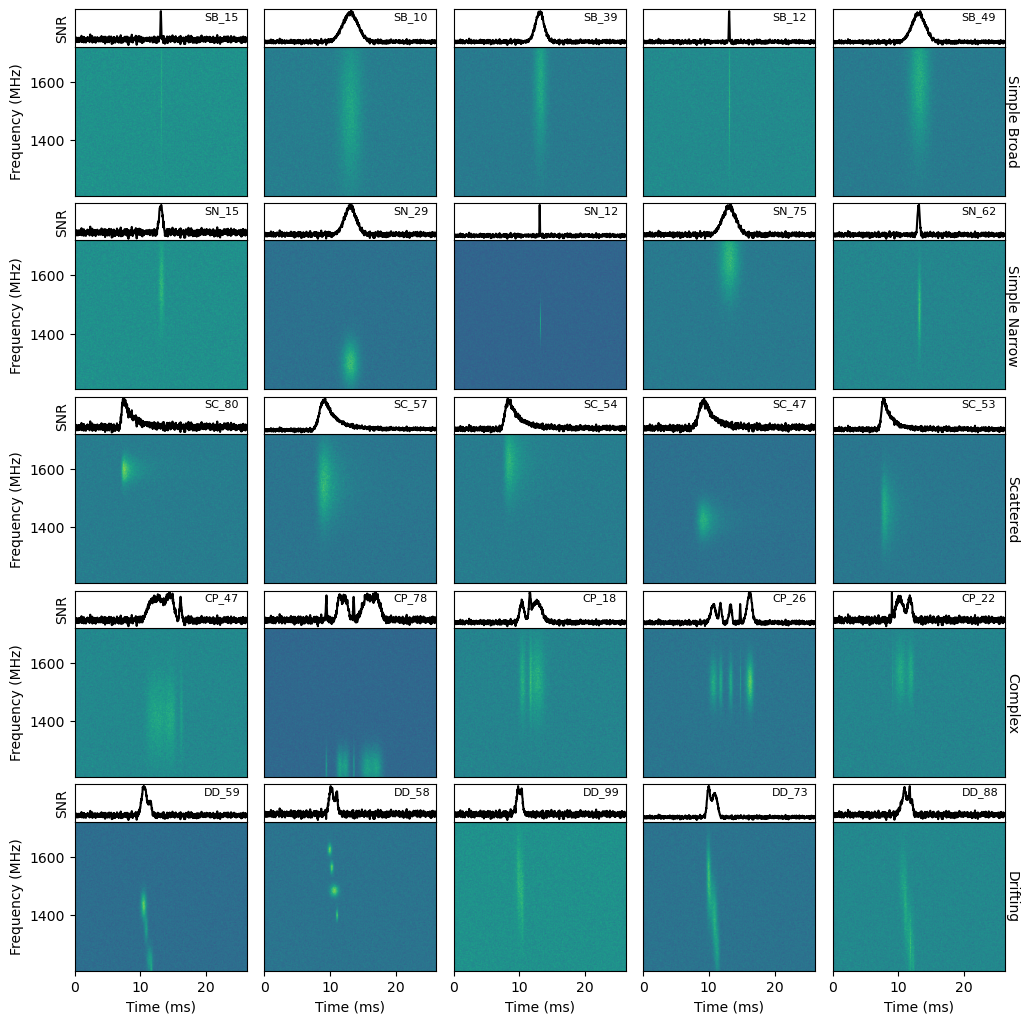}
        \caption{Dynamic spectra with corresponding lightcurves (averaged across all frequencies), for selected bursts that represent different morphological classes. Each row represents a burst category (Simple Broad, Simple Narrow, Scattered, Complex, and Drifting, respectively). Each column shows individual burst examples within that category. Above each dynamic spectrum, the black line represents the lightcurve of the burst. The burst IDs are displayed in the top right of each plot.}
        \label{fig:dynamic_spectra}
    \label{fig:burst_representation}
\end{figure*}

\texttt{FRBakery} provides control over parameters like signal duration, frequency width, and scattering effects, allowing for systematic exploration of how these factors influence representation learning. Figure~\ref{fig:burst_representation} displays 25 simulated bursts from different categories, illustrating their dynamic spectra and light curves.

The simulation generates five main types of FRB datasets with signal-to-noise ratios (S/N) following a power-law distribution:
\begin{itemize}
    \item \textbf{Simple Narrow Bursts:} Gaussian-shaped bursts, both in time and frequency, with varying center frequencies, limited to part of the frequency band.
    \item \textbf{Simple Broad Bursts:} Gaussian-shaped bursts, both in time and frequency, with varying center frequencies, spanning most or all of the frequency band.
    \item \textbf{Scattered Bursts:} Introduce scattering effects, where scattering times vary from $0-20$\,ms.
    \item \textbf{Complex Bursts:} Multi-component bursts with varying temporal and frequency widths for each component.
    \item \textbf{Drifting Bursts:} Frequency-drifting bursts with randomized components and a drift rate between $100-200$\,MHz/ms, resembling the `sad trombone' behavior observed by, e.g., \citet{Hessels_2019}.
\end{itemize}

Each burst category can be generated with adjustable resolutions, depending on the intended application. For initial testing of methods, bursts were typically simulated at a resolution of 1024 by 512 time and frequency bins. For the training set, where simulated data needed to be validated against CHIME observations, bursts were generated at a resolution of 976 by 1024 to match the CHIME dataset.

We simulated a total of 5,000 bursts, evenly distributed across the five morphological categories (Simple Broad, Simple Narrow, Scattered, Complex, and Drifting), with 1,000 bursts per category. This even distribution was chosen to ensure sufficient coverage of all burst types, particularly the Scattered, Complex, and Drifting categories, which exhibit more diverse and challenging features for machine learning models. Table~\ref{tab:dataset_overview} provides a detailed summary of the simulated dataset, including the number of bursts per category, parameter ranges, and distributions. The parameters for each burst were drawn from distributions informed by observed trends and inferred properties of the FRB population, such as signal-to-noise ratio (S/N), frequency drift rate, and scattering time. For the burst time width parameter, the simulations used a Gaussian distribution. The mean of the Gaussian was itself a random variable sampled from a normal distribution. The variance of the Gaussian was drawn from a uniform distribution between the ranges given in Table~\ref{tab:dataset_overview}, resulting in a broadly sampled parameter space.


\begin{table*}
\centering
\renewcommand{\arraystretch}{1.2}
\begin{tabular}
{|p{2.3cm}|p{1.8cm}|p{2.5cm}|p{2.5cm}|p{2.5cm}|p{2.8cm}|}
\hline
\textbf{Dataset Type} & \textbf{Number of Pulses} & \textbf{Time Width (s)} & \textbf{Frequency Width (MHz)} & \textbf{Center Frequency (MHz)} & \textbf{Additional Parameters} \\ \hline \hline
\textbf{Simple Narrow (SN)} & 1000 & $0.0008 \pm 0.0004$ & 12.5--100 & 1228--1700 & S/N scaling exponent = -1.5 \\ \hline
\textbf{Simple Broad (SB)} & 1000 & $0.0008 \pm 0.0004$ & 150--200 & 1300--1625 & S/N scaling exponentt = -1.5 \\ \hline
\textbf{Scattered (SC)} & 1000 & $0.0004 \pm 0.0002$ & 12.5--100 & 1228--1700 & S/N scaling exponent = -1.5,  Scattering Time (\texttt{tau}): $0.001 \pm 0.001$ s \\ \hline
\textbf{Complex (CP)} & 1000 & $0.0004 \pm 0.0002$ & 12.5--100 & 1228--1700 & S/N scaling exponent = -1.5 \\ \hline
\textbf{Drifting (DD)} & 1000 & $0.0002 \pm 0.0001$ & 12.5--100 & 1228--1700 & S/N scaling exponent = -1.5, Drifting Rate = 100--200 MHz/ms.  \\ \hline
\end{tabular}
\caption{Summary of the parameters used for the simulated FRB dataset in FRBakery.}
\label{tab:dataset_overview}
\end{table*}

While the synthetic dataset ensures even coverage of morphological classes for algorithm testing, it is not necessarily representative of the true FRB population. In real-world datasets, some morphologies may dominate while others are underrepresented. Such imbalances can influence the resulting representation space, potentially biasing machine learning algorithms toward overrepresenting certain classes. This is particularly relevant for unsupervised learning algorithms, where the input data distribution significantly shapes the learned representations. Future work could explore how real-world class imbalances impact representation learning and algorithm performance. The synthetic datasets currently do not include the kind of complex burst morphologies observed at the highest signal-to-noise ratios (S/N) and time resolution data \citep[e.g.,][]{faber2023morphologiesbrightcomplexfast, microshots}.

\subsection{CHIME Complex Voltage Data}
\label{sec:chime_baseband}

To validate the simulation framework and test our models on real-world data, we used complex voltage data from the Canadian Hydrogen Intensity Mapping Experiment (CHIME) FRB project. The CHIME/FRB catalog provides a relatively large, uniform dataset, which includes 140 FRBs with available channelized complex voltage data. These data allow for beamforming and coherent dedispersion, offering detailed insights into FRB structures at high time and frequency resolution.

The preprocessing steps for the CHIME data are as follows:
\begin{enumerate}
    \item \textbf{Dimensionality Alignment and Downsampling:} FRBs vary in duration, but machine learning algorithms often require uniform dimensions. To address this, the data was downsampled by a factor of 2 to 32 to reduce the size while preserving the burst structure. After downsampling, all spectra were adjusted to a uniform size of 976 by 1024 (time by frequency bins). For shorter bursts, missing time bins were padded with Gaussian noise that matched the statistical properties of neighboring bins. Longer bursts were truncated to fit the target dimensions.
    \item \textbf{RFI Excision:} radio frequency interference (RFI) was removed iteratively, and missing channels were filled with Gaussian noise.
    \item \textbf{Dedispersion:} Data were incoherently dedispersed using DMs from the CHIME catalog to correct for frequency-dependent time delays.
    \item \textbf{Standardization:} The intensity values of each dynamic spectrum were standardized to have a mean of zero and a standard deviation of one. This was done by subtracting the mean and dividing by the standard deviation of the non-masked intensity values for each burst. Standardization was applied after masking, dedispersion, and dimensionality alignment to ensure uniformity across the dataset.  
\end{enumerate}

These preprocessing steps ensured that the CHIME data was compatible with machine learning pipelines and could be directly compared to the synthetic data generated by \texttt{FRBakery}.

\begin{figure}
    \centering
    \includegraphics[width=\columnwidth]{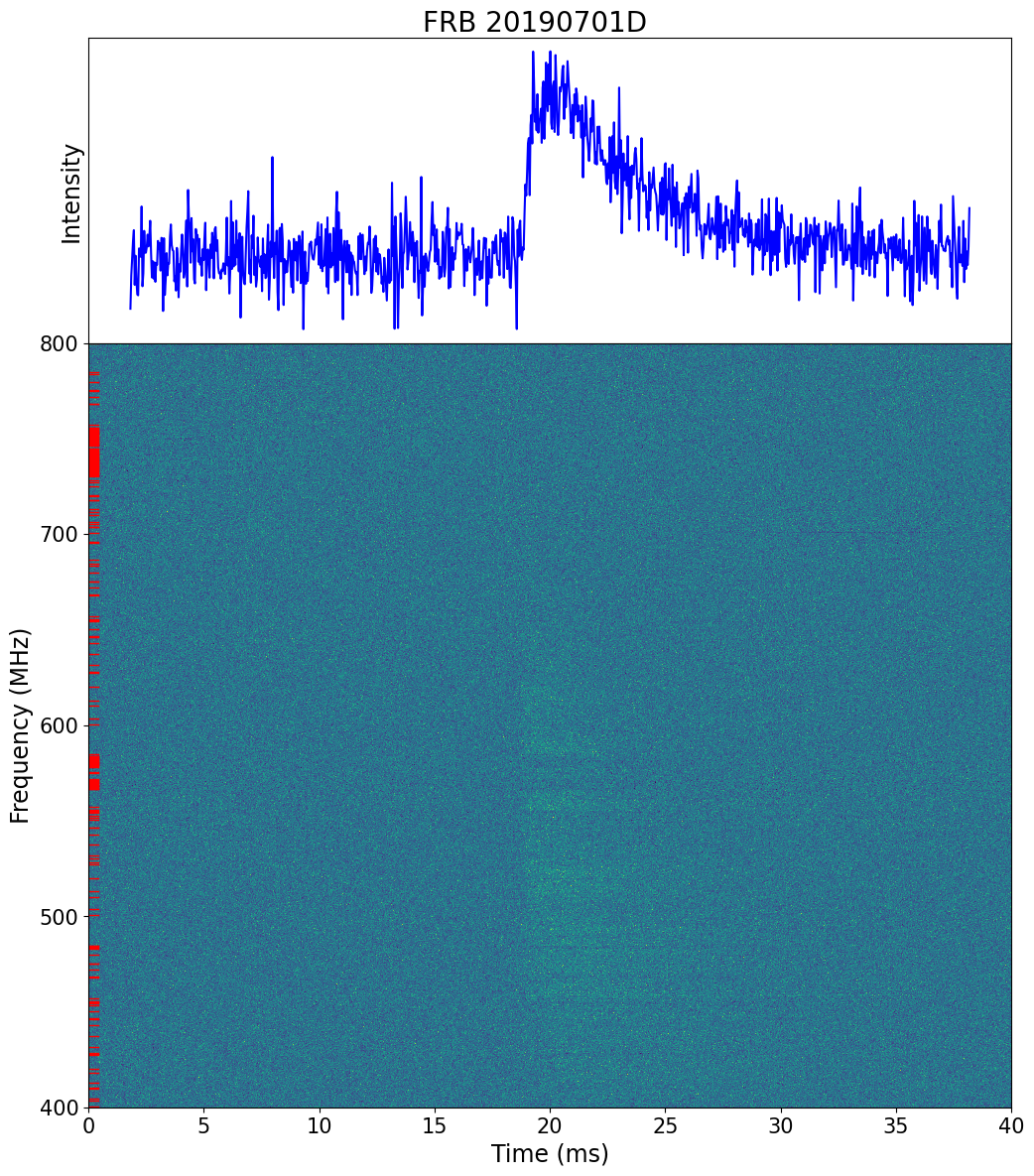}
    \caption{Dynamic spectrum of processed complex voltage data from the CHIME telescope plotting the total intensity (Stokes I). The bottom panel shows the dynamic spectrum of the FRB signal, where each row represents a frequency channel over time. Zapped channels, filled with characteristic noise, are marked with red ticks on the y-axis. The top panel displays the integrated intensity time series, summing the signal across all frequency channels that were not originally zapped.}
    \label{fig:chime_baseband_spectrum}
\end{figure}

Figure~\ref{fig:chime_baseband_spectrum} shows an example of processed complex voltage data from a CHIME burst.

\section{Algorithms and Parameters}
\label{sec: algos}

In this study, we employ two distinct methods for dimensionality reduction and feature extraction: Principal Component Analysis (PCA) and a Convolutional Autoencoder (CAE) enhanced with an Information-Ordered Bottleneck (IOB) layer.

\subsection{Principal Component Analysis}
\label{sec: pca}

PCA is a widely used dimensionality reduction technique that transforms high-dimensional data into a new coordinate system defined by orthogonal axes, known as principal components \citep{jolliffe2002principal}. These components are ordered such that each successive component captures the maximum remaining variance in the data, making PCA a powerful tool for identifying dominant patterns. The outputs of PCA include:
\begin{itemize}
    \item \textbf{Principal Components (PCs):} Linear combinations of the original features that define the new coordinate system.
    \item \textbf{Explained Variance:} The proportion of the total variance captured by each principal component, which helps determine how many components are needed to effectively represent the data.
\end{itemize}

In essence, each principal component corresponds to a pattern of variability, while the associated explained variance quantifies its significance. For example, in an astronomical context, PCA can be used to identify dominant modes of variability in dynamic spectra or other high-dimensional datasets. This method has proven valuable in various astronomical applications, including galaxy morphology classification \citep{pca_galaxies}, exoplanet atmosphere analysis \citep{Damiano2019APC}, and spectroscopic imaging of the interstellar medium \citep{Heyer1997ApplicationOP}.

Given a set of observed \(d\)-dimensional data vectors \(\{\mathbf{x}_n\}\), the principal axes \(\mathbf{W}_j\) are derived from the dominant eigenvectors of the sample covariance matrix:
\[
    \mathbf{S} = \frac{1}{N}\sum_{n=1}^{N}(\mathbf{x}_n - \bar{\mathbf{x}})(\mathbf{x}_n - \bar{\mathbf{x}})^T,
\]
where \(\bar{\mathbf{x}}\) is the sample mean. These eigenvectors \(\mathbf{W}_j\) satisfy:
\[
    \mathbf{S}\mathbf{W}_j = \lambda_j \mathbf{W}_j,
\]
where \(\lambda_j\) represents the variance explained by the \(j\)-th principal component.

The projection of \(\mathbf{x}_n\) onto these principal axes yields a reduced representation:
\[
    \mathbf{z}_n = \mathbf{W}^T(\mathbf{x}_n - \bar{\mathbf{x}}),
\]
where \(\mathbf{W} = (\mathbf{W}_1, \ldots, \mathbf{W}_q)\), with \(q < d\) representing the number of retained components. The proportion of variance captured by each component can be quantified as:
\[
    \text{Explained Variance Ratio (EVR)} = \frac{\lambda_j}{\sum_{k=1}^d \lambda_k}.
\]
PCA minimizes the squared reconstruction error, providing an optimal linear reconstruction of the original data:
\[
    \hat{\mathbf{x}}_n = \mathbf{W}\mathbf{z}_n + \bar{\mathbf{x}}.
\]
This reconstruction represents the best approximation of the original data \(\mathbf{x}_n\) using only the selected \(q\) principal components.
\\
\\
For our application of PCA to FRB dynamic spectra, it is important to note that these spectra are inherently two-dimensional (time and frequency). To apply PCA, we first flattened each 2D spectrum into a 1D vector by concatenating the rows of the spectrum. This is necessary to transform the data into a form suitable for PCA, but it results in the loss of spatial relationships within the original 2D structure. This trade-off is an inherent limitation of using PCA in this context but allows for a systematic exploration of the dominant modes of variability in the data.

\subsection{Convolutional Autoencoder with Information-Ordered Bottleneck (IOB)}

Neural networks are powerful computational tools inspired by the way biological neurons process information \citep{lecun2015}. They are particularly well-suited to finding patterns and relationships in complex datasets, like FRB dynamic spectra, by learning directly from the data itself without requiring explicitly programmed rules. Within this framework, autoencoders are a specific type of neural network designed to compress data into a compact, lower-dimensional representation (called the `latent space') and then reconstruct the original data from this compressed form \citep{hinton2006}. For FRB dynamic spectra, this means representing the complex time-frequency structure in a way that retains essential information while discarding noise and redundancy.

\subsubsection{Autoencoders}
An autoencoder consists of two main components: an encoder and a decoder. The encoder compresses the input data into a smaller latent representation, while the decoder reconstructs the data from this compressed representation. This process forces the network to learn the most critical features of the input. Mathematically, the encoder \(e_\phi\) maps the input data \(x \in \mathcal{X}\) to a lower-dimensional latent representation \(z \in \mathcal{Z}\), while the decoder \(d_\eta\) reconstructs the data back into the original space \(\mathcal{X}\). This can be expressed as:
\[
e_\phi: \mathcal{X} \to \mathcal{Z}, \quad z = e_\phi(x),
\]
\[
d_\eta: \mathcal{Z} \to \mathcal{X}, \quad \hat{x} = d_\eta(z),
\]
where \(\phi\) and \(\eta\) are the learnable parameters of the encoder and decoder, respectively, and \(\hat{x}\) is the reconstructed input. 

The autoencoder is trained by minimizing a reconstruction loss, which measures the difference between the original input \(x\) and its reconstruction \(\hat{x}\). The most common reconstruction loss is the Mean Squared Error (MSE):
\[
\mathcal{L}_{\text{reconstruction}} = \frac{1}{N} \sum_{i=1}^N \|x_i - \hat{x}_i\|^2,
\]
where \(N\) is the number of elements in the dataset.

The key idea of an autoencoder is to pass the input through a bottleneck layer (the latent space) where the network is forced to compress the information into a much smaller number of variables. This encourages the network to retain only the most important features of the input while discarding noise and less relevant information.

\subsubsection{Convolutional Autoencoders}
A CAE extends this concept by using convolutional layers in the encoder and decoder, which are particularly well-suited for structured data like images or dynamic spectra \citep{Masci2011StackedCA}. Convolutional layers apply filters across the input to detect patterns, such as edges, textures, or frequency trends, at different spatial scales. These filters allow the network to efficiently capture spatial dependencies in the data, which is convenient for FRB dynamic spectra because they contain both time and frequency information and can effectively be treated as an image. 

In the encoder, convolutional layers progressively reduce the spatial dimensions of the input data while increasing the feature depth, compressing the data into a lower-dimensional latent representation. In the decoder, transposed convolutional layers perform the opposite operation, reconstructing the data back to its original dimensions.

The goal of the CAE is to compress the dynamic spectra into a small set of numbers (the latent variables) in a way that retains the essential information needed to reconstruct the original input. This is particularly valuable for FRB analysis because it allows us to represent complex bursts in a compact, interpretable format. By feeding the dynamic spectra through this bottleneck, the CAE can more effectively separate important features from noise and redundancies.

Autoencoders and CAEs are commonly used for tasks like noise removal, feature extraction, and dimensionality reduction, making them a natural choice for analyzing the high-dimensional, structured data associated with FRBs \citep{Masci2011StackedCA, Krizhevsky2012ImageNetCW, lecun2015}. However, traditional autoencoders often produce latent spaces that are unstructured and difficult to interpret, which is why we enhance our CAE with the IOB layer.

\subsubsection{Information-Ordered Bottleneck}
\label{sec: iob}
The IOB addresses these limitations by adaptively compressing data and prioritizing the most important features in the latent space. This is achieved by dynamically varying the size of the bottleneck during training and ordering the latent variables by their contribution to the reconstruction.

In the IOB framework, an additional bottleneck function \(b_k\) is introduced, which masks the latent space \(z \in \mathcal{Z}\) to a dimensionality \(k\) (where \(k \leq \dim(\mathcal{Z})\)). This creates an adjustable latent space during training. For a bottleneck of size \(k\), the mapping becomes:
\[
f_\theta^{(k)}(x) = d_\eta ( b_k (\ e_\phi(x))),
\]
where \(f_\theta^{(k)}\) is the model with a bottleneck width of \(k\), and \(\theta = \{\phi, \eta\}\) are the learnable parameters of the model.



\subsubsection{Conceptual Design of IOB}

\begin{figure}
    \centering
    \includegraphics[width=\columnwidth]{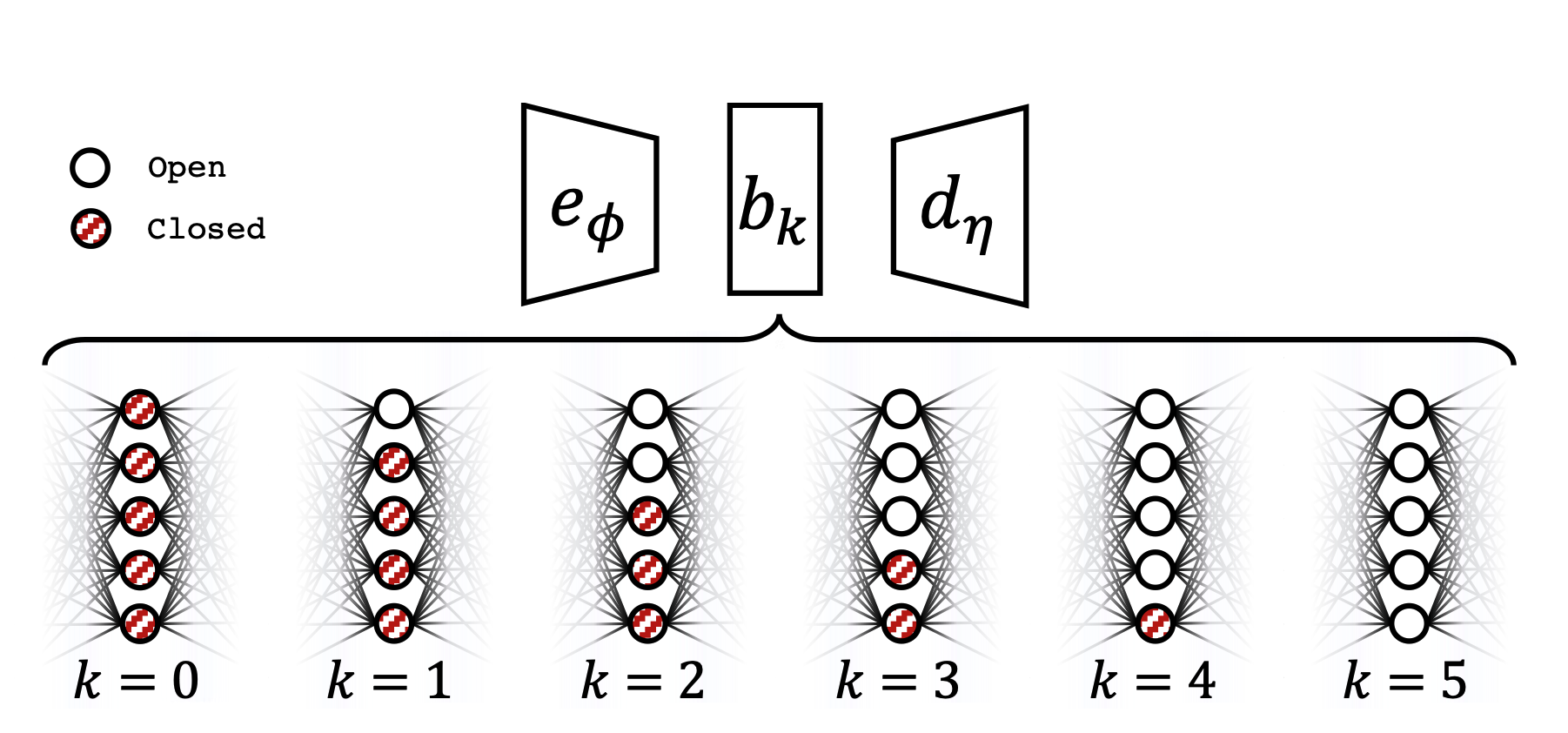}
    \caption{Conceptual design of the IOB. At each training step, the bottleneck width \(k\) is varied by masking inactive latent variables. The active variables are prioritized by their contribution to the reconstruction, creating a structured and ordered latent space. Figure taken from \citet{ho2023informationordered}.}
    \label{fig:IOB_concept}
\end{figure}

Figure~\ref{fig:IOB_concept} illustrates the conceptual design of the IOB. During training, the bottleneck width \(k\) is incrementally varied, with only the first \(k\) latent variables being active. Latent variables beyond this width are masked, meaning they contribute no information and do not propagate gradients. This encourages the network to maximize the information passed through the top latent variables, which are open and active. As training progresses, this adaptive mechanism orders the latent variables by their contribution to the reconstruction, resulting in a compact and interpretable latent space.

This iterative process ensures that the most reliable pathways are used for reconstruction, allowing the IOB to create an efficient representation of the data. The resulting latent space is both structured and compact, where the top latent variables capture the most critical information about the input data.

\begin{figure*}
    \centering
    \includegraphics[width=\textwidth]{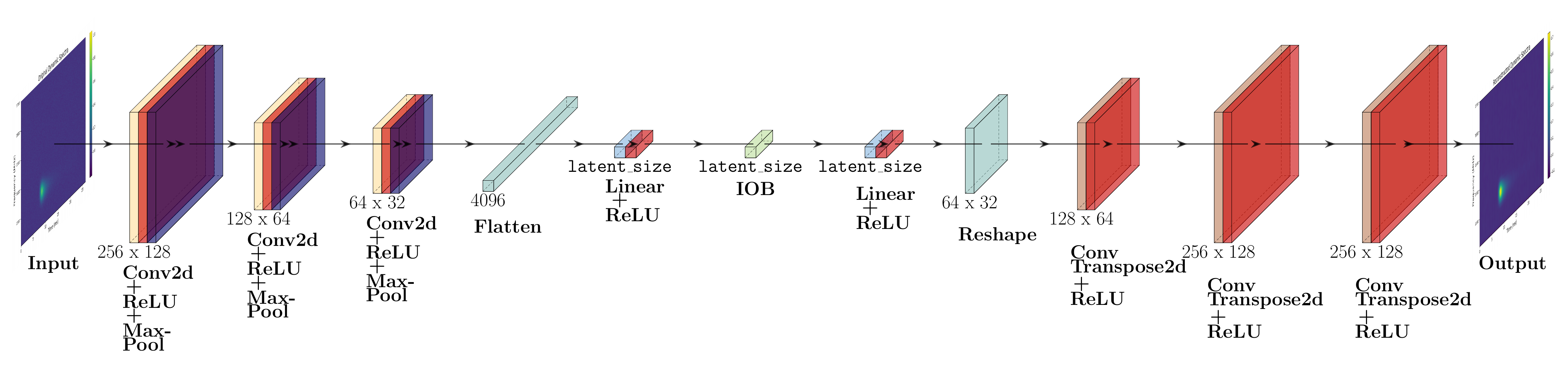}
    \caption{Structure of the Convolutional Autoencoder (CAE) with the Information Ordered Bottleneck (IOB) layer. The encoder compresses the input data, the IOB layer refines the latent representation by adaptively ordering latent variables, and the decoder reconstructs the input from the refined latent space.}
    \label{fig:CAE_IOB_structure}
\end{figure*}

\subsubsection{Structure of the CAE with IOB}

The architecture of the CAE with the IOB layer is depicted in Figure~\ref{fig:CAE_IOB_structure}. The CAE consists of three main components: an encoder, the IOB layer, and a decoder. Each component was carefully designed to effectively process and compress the dynamic spectra while preserving essential spatial and temporal features.

\paragraph*{Encoder:}
The encoder processes the input dynamic spectrum, which was initially downsampled to a resolution of \(256 \times 128\) (time-frequency), and progressively compresses it into a lower-dimensional latent representation. It consists of two convolutional layers, each followed by a Rectified Linear Unit (ReLU) activation function and max-pooling. The convolutional layers reduce the spatial size while increasing feature depth to capture hierarchical patterns in the data.

\begin{itemize}
    \item \textbf{Layer 1:} 
        \begin{itemize}
            \item 2D convolutional layer with:
                \begin{itemize}
                    \item Kernel size: \(3 \times 3\),
                    \item Stride: \(2\),
                    \item Padding: \(1\).
                \end{itemize}
            \item Followed by ReLU activation and a \(2 \times 2\) max-pooling layer.
            \item Output size: \(128 \times 64\).
        \end{itemize}
    \item \textbf{Layer 2:} 
        \begin{itemize}
            \item 2D convolutional layer with:
                \begin{itemize}
                    \item Kernel size: \(3 \times 3\),
                    \item Stride: \(2\),
                    \item Padding: \(1\).
                \end{itemize}
            \item Followed by ReLU activation and a \(2 \times 2\) max-pooling layer.
            \item Output size: \(64 \times 32\).
        \end{itemize}
\end{itemize}

After the second convolutional layer, the output feature map is flattened into a 1D vector of size \(4096\). This vector is then passed through a fully connected (dense) layer with ReLU activation, which reduces it to the latent size required by the IOB layer.

\paragraph*{IOB:}
The IOB layer processes the latent representation produced by the encoder. It adaptively compresses the data by ordering and masking latent variables based on their contribution to reconstruction, as described in Section~\ref{sec: iob}.

\paragraph*{Decoder:}
The decoder reconstructs the original input from the compressed latent representation, mirroring the structure of the encoder. Using transposed convolutional layers, it progressively upsamples the latent space back to the original input dimensions (\(1024 \times 512\)). Specifically, the decoder includes transposed convolutional layers with the same kernel size, stride, and padding as the encoder's convolutional layers but applied in reverse.

\paragraph*{Architecture Selection:}
The CAE architecture was chosen through iterative experimentation. Starting with a simple design, we gradually added complexity by adjusting the number of layers, kernel sizes, and latent space dimensions, monitoring the reconstruction loss. The final configuration strikes a balance between reconstruction accuracy and computational efficiency. We found that deeper architectures did not yield significant improvements and often led to overfitting. The ReLU activation function was selected due to its computational efficiency and widespread success in convolutional networks for similar applications.

\begin{figure*}
    \centering
    \begin{subfigure}[t]{0.48\textwidth}
        \centering
        \includegraphics[width=\textwidth]{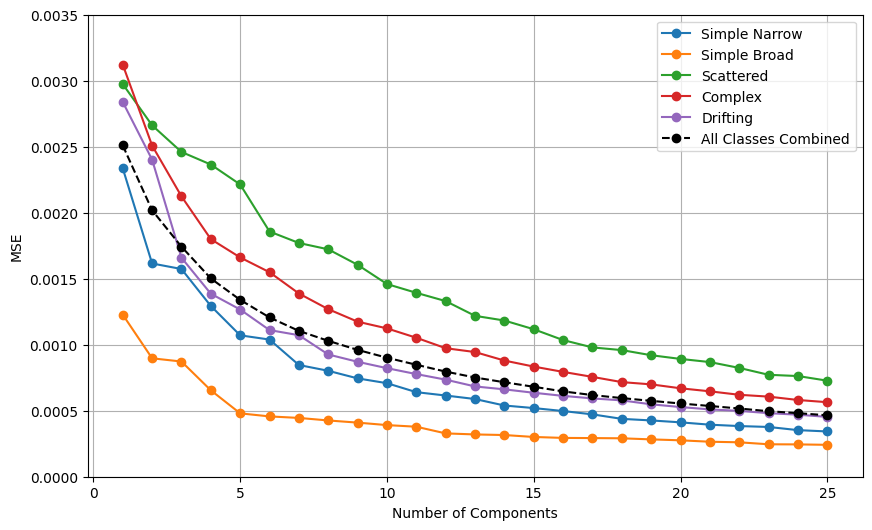}
        \caption{}
        \label{fig:pca_mse}
    \end{subfigure}
    \hfill
    \begin{subfigure}[t]{0.48\textwidth}
        \centering
        \includegraphics[width=\textwidth]{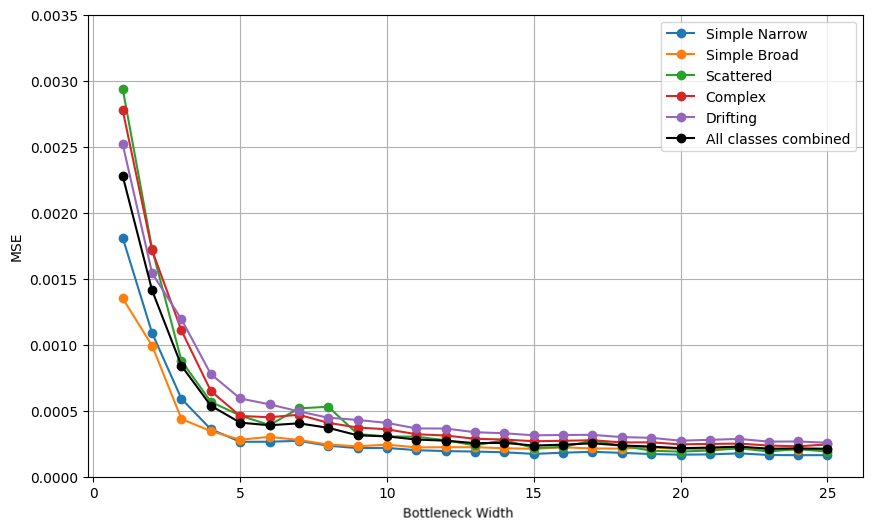}
        \caption{}
        \label{fig:mse_iob_conv}
    \end{subfigure}
    \caption{Comparison of reconstruction performance for PCA and IOB-CAE. The left panel shows the MSE as a function of the number of PCA components, with decreasing MSE indicating improved reconstruction accuracy. The right panel illustrates MSE as a function of bottleneck width (number of latent variables) for the IOB-CAE, showing how the reconstruction error stabilizes as the bottleneck width increases.}
    \label{fig:reconstruction_comparison}
\end{figure*}

\subsubsection{Training the CAE with IOB}

The training process aims to optimize the parameters of the CAE with the IOB to ensure the model learns how to accurately compress and reconstruct dynamic spectra. This is achieved by minimizing a reconstruction loss, which quantifies the difference between the original input spectrum and its reconstruction. The Mean Squared Error (MSE) is used as the loss function.

To optimize the model, we use the Adam optimizer \citep{kingma2017adam}, a commonly used algorithm in machine learning that adjusts the learning rate for each parameter dynamically.

To prevent overfitting, we employ an early stopping mechanism. Early stopping monitors the reconstruction loss on a test dataset that is not used during training. If the test loss does not improve for 20 consecutive epochs, training is halted. This approach helps the model generalize well to unseen data, ensuring that it learns meaningful patterns rather than memorizing noise or redundant features \citep{Prechelt1996EarlySW, Ying2019AnOO}.

By combining these techniques, the CAE with IOB is trained to produce a structured and interpretable latent space. This latent space organizes the most critical features of the input spectra, enabling further analysis, such as clustering or visualizing FRB morphologies, while retaining essential information from the original data.

\begin{figure*}
    \centering
    \begin{subfigure}[t]{0.48\textwidth}
        \centering
        \includegraphics[width=\textwidth]{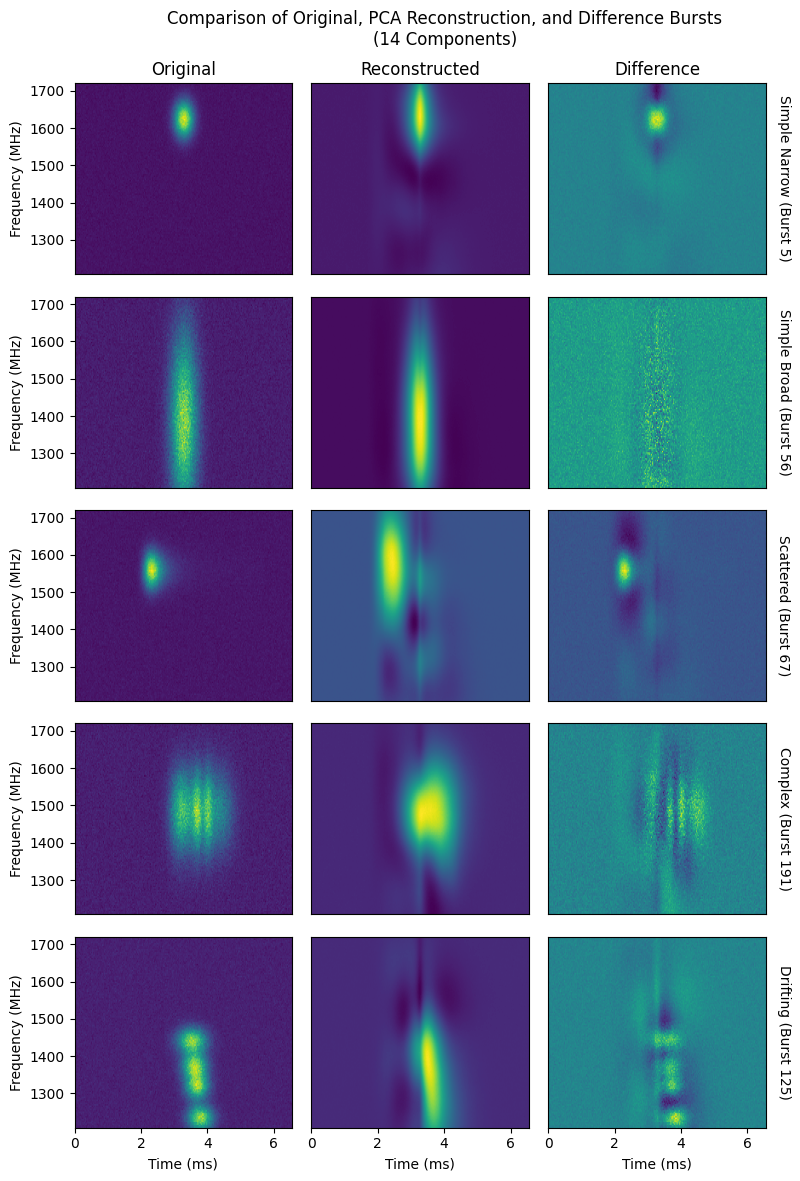}
        \caption{}
        \label{fig:reconstruction_pca}
    \end{subfigure}
    \hfill
    \begin{subfigure}[t]{0.48\textwidth}
        \centering
        \includegraphics[width=\textwidth]{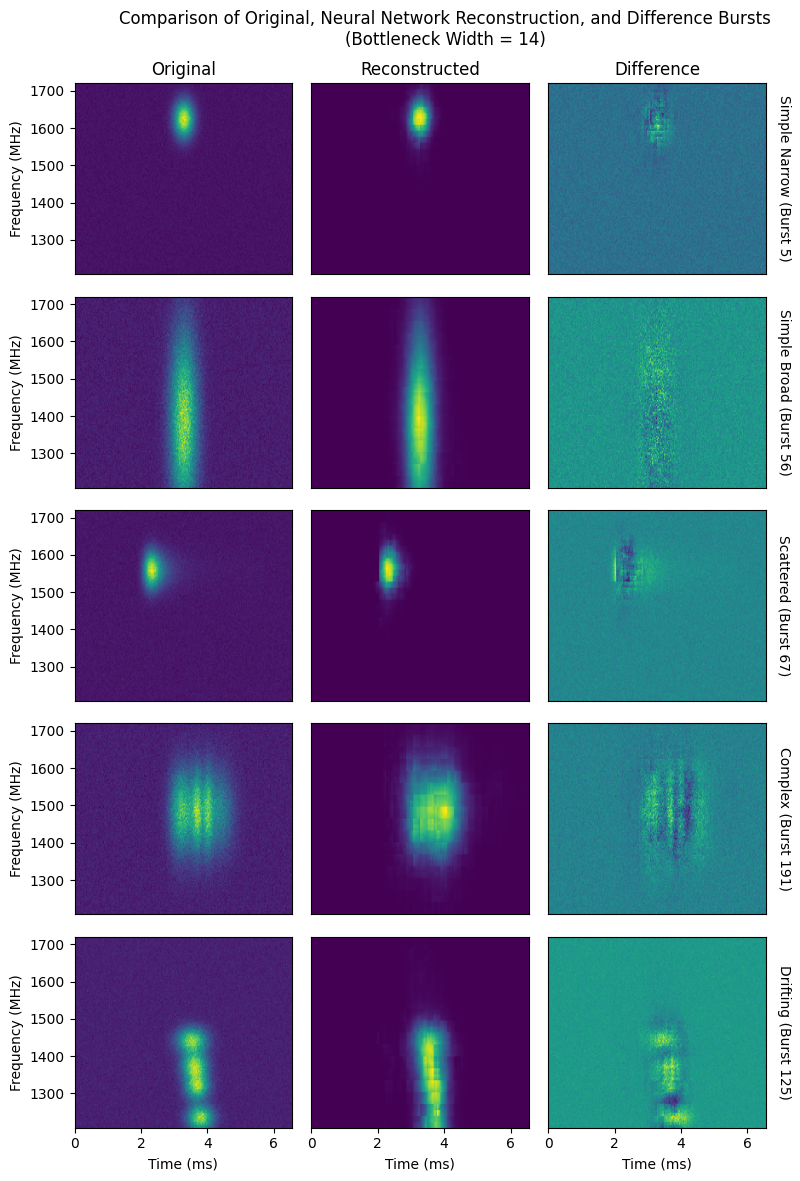}
        \caption{}
        \label{fig:reconstruction_iob}
    \end{subfigure}
    \caption{Comparison of PCA and IOB-CAE reconstructions for different FRB burst classes (Simple Broad, Simple Narrow, Scattered, Complex, and Drifting). Each row corresponds to a burst category, with three columns displaying the original burst, the reconstructed burst, and the difference between the original and reconstructed bursts. For PCA (left panel), reconstructions were performed using 14 components, as the explained variance plateaued at this value. For IOB-CAE (right panel), reconstructions were performed using a bottleneck width of 14, providing a comparable dimensionality for both methods.}
    \label{fig:reconstruction_comparison}
\end{figure*}

\section{Simulation-Based Analysis} \label{sec: results1}

\subsection*{Reconstruction Performance and Quality} \label{sec: rec_performance}

We applied both PCA and the IOB-CAE to the simulated dynamic spectra to compare their performance in reconstructing FRB signals. To evaluate reconstruction performance, we calculated the mean squared error (MSE) as a function of the number of components or latent variables used.

The MSE was chosen as the evaluation metric because it quantifies the average squared difference between the original and reconstructed data, providing a direct measure of reconstruction accuracy. A lower MSE indicates that the reconstructed data closely resembles the original input, reflecting the model’s ability to capture and retain essential features of the FRB signals. Conversely, a higher MSE suggests that important features were not adequately preserved, leading to poor reconstruction.

This metric is particularly useful for comparing dimensionality reduction methods, such as PCA and the IOB-CAE, as it allows us to assess how effectively each method compresses and reconstructs data. By observing the behavior of MSE as a function of the number of components (for PCA) or latent variables (for the IOB-CAE), we can evaluate how efficiently each method balances compression and reconstruction quality. For example, we expect the MSE to decrease as more components or latent variables are added, since the models should capture increasing amounts of information. However, a plateau in MSE indicates that adding more components or variables no longer provides significant improvements, suggesting an optimal number for reconstructin or that the model has reached a limit in complexity that it can effectively reconstruct..

The dataset was divided into an 80/20 split for training and testing. For the IOB-CAE, the model was trained on the 80\% training set, and its performance was assessed on the 20\% test set. For PCA, the principal components were derived from the training set and then used to reconstruct the test set for evaluation. This approach ensures a fair comparison between the methods by assessing their ability to generalize to unseen data.

Figure~\ref{fig:pca_mse} shows the MSE as a function of the number of PCA components. For PCA, the MSE steadily decreases with an increasing number of components. The most significant improvements occur with the first few components, which capture the bulk of the variance in the data, explaining the broad structure of the dynamic spectra. As more components are added, the reduction in MSE levels off. This behavior is expected, as subsequent components account for progressively smaller-scale details. The plateau in MSE indicates that PCA’s ability to improve reconstruction diminishes with additional components.

Despite the steady decline in MSE shown in Figure~\ref{fig:pca_mse}, PCA struggles to reconstruct more complex bursts, such as scattered, complex, and drifting types. The three lines with the highest MSE in the plot correspond to these more intricate burst classes, emphasizing PCA’s difficulty in encoding their detailed structures. This limitation arises because PCA captures variance in a linear manner, which works well for simpler signals but is insufficient for the non-linear patterns characteristic of complex FRB bursts. 

For example, while the MSE for simpler classes, such as Simple Broad and Simple Narrow, drops quickly with the addition of components, the more complicated bursts show a much slower decline in MSE and remain significantly higher overall. This indicates that even as more components are added, PCA fails to adequately capture the intricate features of these complex bursts. Most of the initial components in PCA focus on global variance, which is often dominated by noise. As a result, the detailed signal information necessary to distinguish between burst types is either underrepresented or lost entirely.

These limitations are further reflected in the reconstruction plots in Figure~\ref{fig:reconstruction_pca}. This figure provides a qualitative comparison of the original bursts and their reconstructions using PCA with 14 components, the point at which the MSE begins to flatten. PCA effectively reconstructs simpler bursts, such as Simple Broad and Simple Narrow, capturing their widths in time and frequency band positions. However, as burst complexity increases—for example, in Scattered and Complex types (rows 3 and 4)—the reconstructions deviate significantly from the originals. The PCA reconstructions for complex bursts lack the finer details needed to fully represent their structures and show artifacts in the reconstructions, underscoring the limitations of a linear dimensionality reduction method. This qualitative assessment aligns with the quantitative trends observed in Figure~\ref{fig:pca_mse}, where the MSE for scattered, complex and drifting bursts remains significantly higher than that for simpler bursts. Together, these results show that while PCA is sufficient for reconstructing broad, simple features, it fails to preserve the intricate, non-linear characteristics of more complicated FRB signals, making it less effective for analyzing their full complexity.

The IOB-CAE, in contrast, displayed significantly better performance in reconstructing FRB bursts, particularly the more complex ones. Figure~\ref{fig:mse_iob_conv} shows the MSE as a function of the latent bottleneck width. Similar to PCA, the MSE steadily decreases as the number of latent variables increases, but the trend is strikingly different. The IOB-CAE shows a much steeper initial decline in MSE, indicating that it captures the essential features of the input data more efficiently, particularly for complex bursts. This trend contrasts sharply with the slower and more gradual decline seen in Figure~\ref{fig:pca_mse} for PCA, where more components are needed to achieve similar levels of reconstruction quality.

The plateau in the MSE for the IOB-CAE, which occurs at around 6-8 latent components, suggests that this small number is sufficient to encode both global structures and fine-grained details of the FRB bursts. In comparison, PCA requires significantly more components to achieve a comparable MSE, reflecting its limitations in handling non-linear dependencies. This steep decline and early stabilization of the MSE in the IOB-CAE highlight its expressive power and efficiency in compressing information, even for complex and scattered bursts.

To support these observations, it is crucial to compare the reconstruction plots directly. As shown in Figure~\ref{fig:reconstruction_iob}, the IOB-CAE reconstructions preserve both the global structure and intricate features of complex bursts, such as drifting and scattered types, which are more accurately represented than in the PCA reconstructions (Figure~\ref{fig:reconstruction_pca}). This qualitative improvement aligns with the quantitative trends: while PCA struggles to capture the fine details necessary for reconstructing these burst types, the IOB-CAE demonstrates its capacity to handle these challenging features more effectively.

An important consideration here is the role of noise in the MSE values. For  many spectra in the dataset, the bins containing only measurement noise outnumber the bins containing FRB signal by over an order of magnitude. Since the MSE sums deviations across all bins, algorithms that are good at reconstructing noise can achieve similar MSE values even if their ability to reconstruct signal differs substantially. This is particularly relevant for PCA, whose linear approach often prioritizes capturing global variance, but may struggle to encode finer burst structures when noise dominates the dataset.
In contrast, the IOB-CAE’s non-linearity allows it to encode a broader range of data properties, including those encapsulating burst structures, leading to better reconstructions overall. Rather than simply optimizing for variance, the IOB-CAE can learn to represent complex features of the bursts more effectively.
Thus, beyond the absolute MSE score of a given model, we are interested in two key aspects: (1) how quickly the MSE decreases as the number of components or latent variables increases, and (2) the differences in MSE trends between different burst classes. The IOB-CAE achieves both objectives more effectively than PCA, as evident in Figure~\ref{fig:mse_iob_conv}. The steeper MSE decline and earlier plateau reflect its superior ability to compress and reconstruct data, while the reconstruction plots confirm its robustness in handling diverse burst complexities.

Together, these results demonstrate that the IOB-CAE not only minimizes MSE more efficiently than PCA but also retains more critical information about complex bursts. Its ability to accurately reconstruct a wide range of burst types underscores its advantage over PCA, particularly for analyzing FRB signals dominated by noise and non-linear features.

\begin{figure*}
    \centering
    \includegraphics[width=\textwidth]{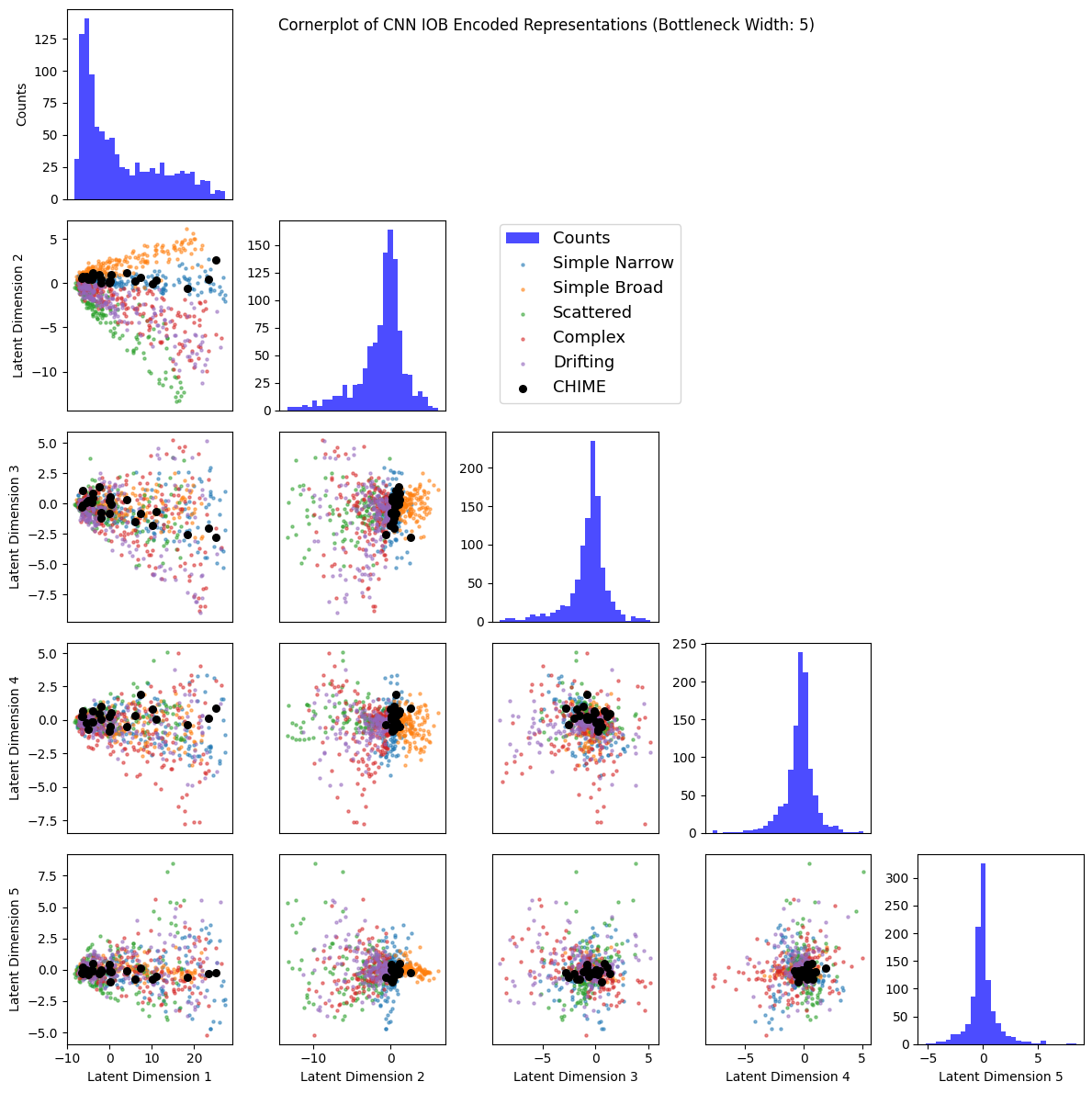}
    \caption{First five components of the latent space for the convolutional autoencoder with IOB using both simulated and real CHIME bursts. Different colors represent different classes.}
    \label{fig:latent_space_components_real}
\end{figure*}

\section{Extending to Real Data}
\label{sec: results2}
\subsection{Latent Space Representation}

To evaluate how well the IOB-CAE and PCA capture the relationships between simulated and real FRB bursts, we trained both methods on a combined dataset of simulated FRBs and real CHIME complex voltage data bursts. The simulated bursts were generated to closely resemble CHIME data by matching dynamic range, noise levels, burst complexity, and data structure. Specifically, we incorporated the effects of RFI zapping observed in the CHIME data, where certain frequency channels containing excessive noise or interference were flagged and excluded from the analysis. These zapped channels, which created gaps in the frequency spectrum, were replicated in the simulations to ensure consistency with the real CHIME data. Additionally, the simulated bursts were adjusted to match the dimensionality of the CHIME data, accounting for the same frequency and time resolutions.

The combined dataset was split into 80\% training data and 20\% test data. Both simulated and real CHIME bursts were included in the training set, ensuring that the model could learn shared features from both types of data. The test set was used to evaluate the generalization ability of both methods to unseen bursts. For PCA, the components were derived from the training set, and reconstructions and latent space projections were performed on the test set. This approach ensured that the training and evaluation processes accurately reflected the real data’s characteristics, allowing for a robust comparison of the methods' performance in capturing shared structures between simulated and real FRB bursts.

\begin{figure*}
    \centering
    \includegraphics[width=\textwidth]{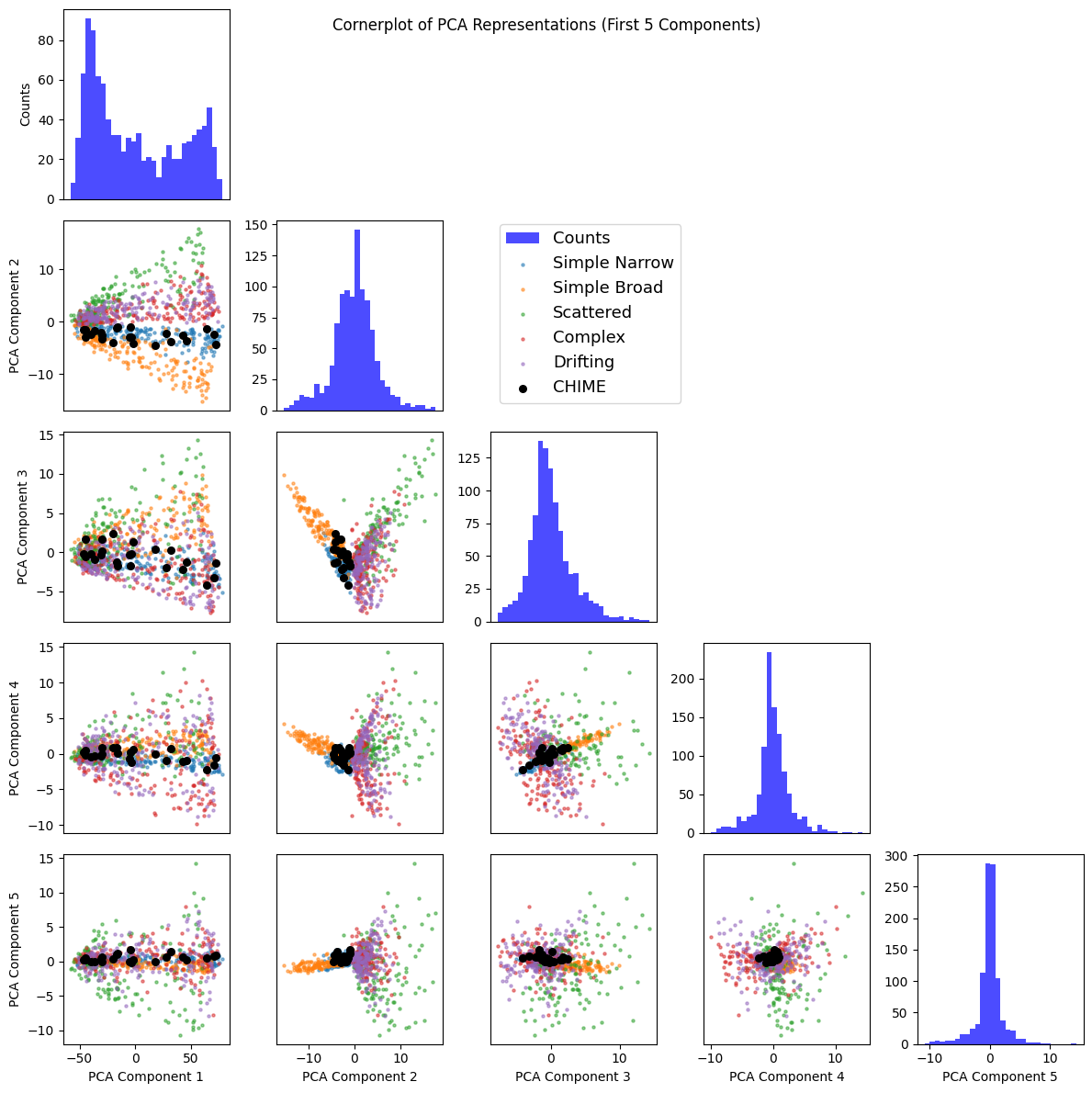}
    \caption{First five components of PCA using both simulated and real CHIME bursts. Different colors represent different classes.}
    \label{fig:latent_space_components_pca}
\end{figure*}

The IOB-CAE latent space representation reveals initial patterns of separation for certain morphology classes, though further refinement is anticipated as additional real bursts are incorporated into the dataset. While the coordinates in the IOB-CAE latent space lack inherent physical meaning, bursts that are closer to one another tend to share more similarities, indicating that the model captures shared features across the dataset. Simpler morphology classes, such as Simple Broad and Simple Narrow, exhibit tighter grouping in this latent space, as seen in Figure~\ref{fig:latent_space_components_real}, while more complex classes like Scattered, Drifting, and Complex tend to overlap, forming a continuum rather than distinct clusters. This aligns with the reconstruction trends observed in Section~\ref{sec: results1}, where simpler bursts were reconstructed more accurately than their complex counterparts.

Notably, the simpler narrow-band CHIME bursts align well with their simulated counterparts in the IOB-CAE latent space (blue points in Figure~\ref{fig:latent_space_components_real}, particularly in dimensions analogous to PC1 and PC2). This suggests that the IOB-CAE effectively captures the essential features required for differentiating these FRBs. As more real bursts are added to the training set, we anticipate further improvements in class separability, especially for more complex bursts where overlap currently remains prevalent.

Interestingly, the PCA latent space (Figure~\ref{fig:latent_space_components_pca}) exhibits slightly more structure compared to the IOB-CAE representation. For example, the narrow-band CHIME bursts (black points) cluster closely with simulated narrow bursts in PC1 vs. PC2 space, suggesting that PCA can identify shared features between real and simulated data in the most significant components. Principal components for simpler classes show tighter clustering, and there is some separation between classes with distinct morphology, particularly in PC3 and PC4. However, much like the IOB-CAE, PCA forms a continuum for the majority of bursts, with many ending up in the denser bulk regions of the space. Without color coding, the boundaries between classes would remain difficult to distinguish. This continuum may reflect both the intrinsic diversity of the FRB population parameters and the nature of the dynamic spectra in pixel space. For example, in pixel space, there can be a natural progression from narrow-band to drifting bursts, as well as a wide range of features from narrow to broad bursts (as illustrated in Figure~\ref{fig:burst_representation}). The distinction between `broad' and `narrow' bursts is inherently somewhat arbitrary, suggesting that observed overlaps in latent space are a natural outcome of this continuum in the data. This continuum may also indicate that the underlying processes governing the FRB population share overlapping characteristics rather than forming discrete clusters.

It is notable that neither method identified complex bursts as clear outliers. Instead, complex bursts often blended into the bulk of the latent space, likely due to their less accurate reconstruction (Section~\ref{sec: results1}). This is evident in the PCA latent space (Figure~\ref{fig:latent_space_components_pca}), where complex bursts overlap significantly with other classes, and in the IOB-CAE space (Figure~\ref{fig:latent_space_components_real}), where the same trend is observed. This suggests that reconstruction quality plays a role in shaping the latent space structure, with simpler, better-reconstructed bursts forming tighter clusters and complex bursts contributing to the continuum.

These results highlight the importance of gathering additional real data to better understand the latent space dynamics. Expanding the dataset size and incorporating more real bursts could reveal whether the observed continuum persists or evolves into more distinct groupings, particularly for complex bursts that currently exhibit significant overlap with other classes.

\subsection{Outlier Detection with PCA}

The analysis of latent spaces not only provides insights into the overall structure and relationships between bursts but can also highlight anomalous patterns that may signify outliers. Building on the observations from the latent space representations, we applied PCA directly to the CHIME dataset, without integrating simulated bursts, to explore whether this dimensionality reduction technique could uncover previously hidden structures and detect bursts with unusual features. By leveraging PCA’s ability to project high-dimensional data into a lower-dimensional space, we aimed to identify outliers based on their deviations in the reduced feature space. A similar approach was not applied using the IOB-CAE, as the CHIME dataset alone provides too little data to effectively train the non-linear model without incorporating simulated bursts, limiting its ability to generalize.

Using an interactive visualization tool\footnote{The code for the visualization tool can be found at: \url{https://github.com/SRON-API-DataMagic/Rep_Learn_FRB/tree/main/FRBakery}}, we explored the principal components obtained from PCA, where each burst is projected onto the first two principal components. This approach enabled an intuitive examination of individual bursts and facilitated the identification of outliers.

\begin{figure*}
    \centering
    \includegraphics[width=\textwidth, trim=0 0 0 70, clip]{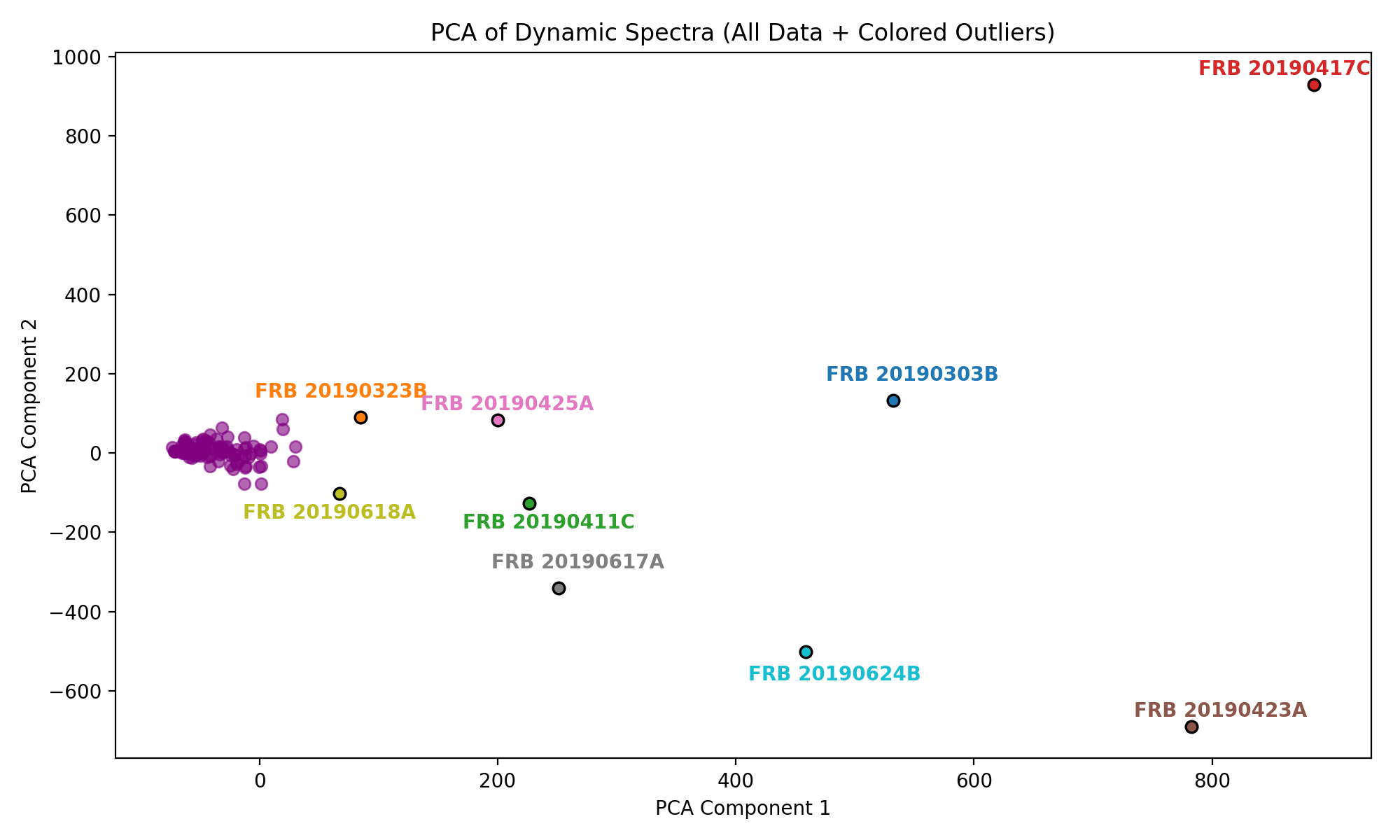}
    \caption{PCA visualization of the CHIME bursts, highlighting multiple outliers. The first two principal components are shown, with the most significant outliers labeled and color-coded for further analysis. Many of these outliers exhibit complex burst structures, such as multiple components or scattering tails, in contrast to the more typical one-component Gaussian bursts that cluster in the low PC1 and PC2 region.}
    \label{fig:pca_analysis1}
\end{figure*}

Figure~\ref{fig:pca_analysis1} shows the results of this analysis, where bursts are projected onto the first two principal components of the reduced feature space. Several bursts stand out as outliers, exhibiting significant deviations in both the first and second principal components (PC1 and PC2), suggesting unusual features in their dynamic spectra. The most prominent outliers include FRB 20190417C, FRB 20190423A, FRB 20190303B, FRB 20190624B, FRB 20190411C, FRB 20190617A, FRB 20190618A, FRB 20190425A, and FRB 20190323B.

\begin{figure*}
    \centering
    \includegraphics[width=\textwidth]{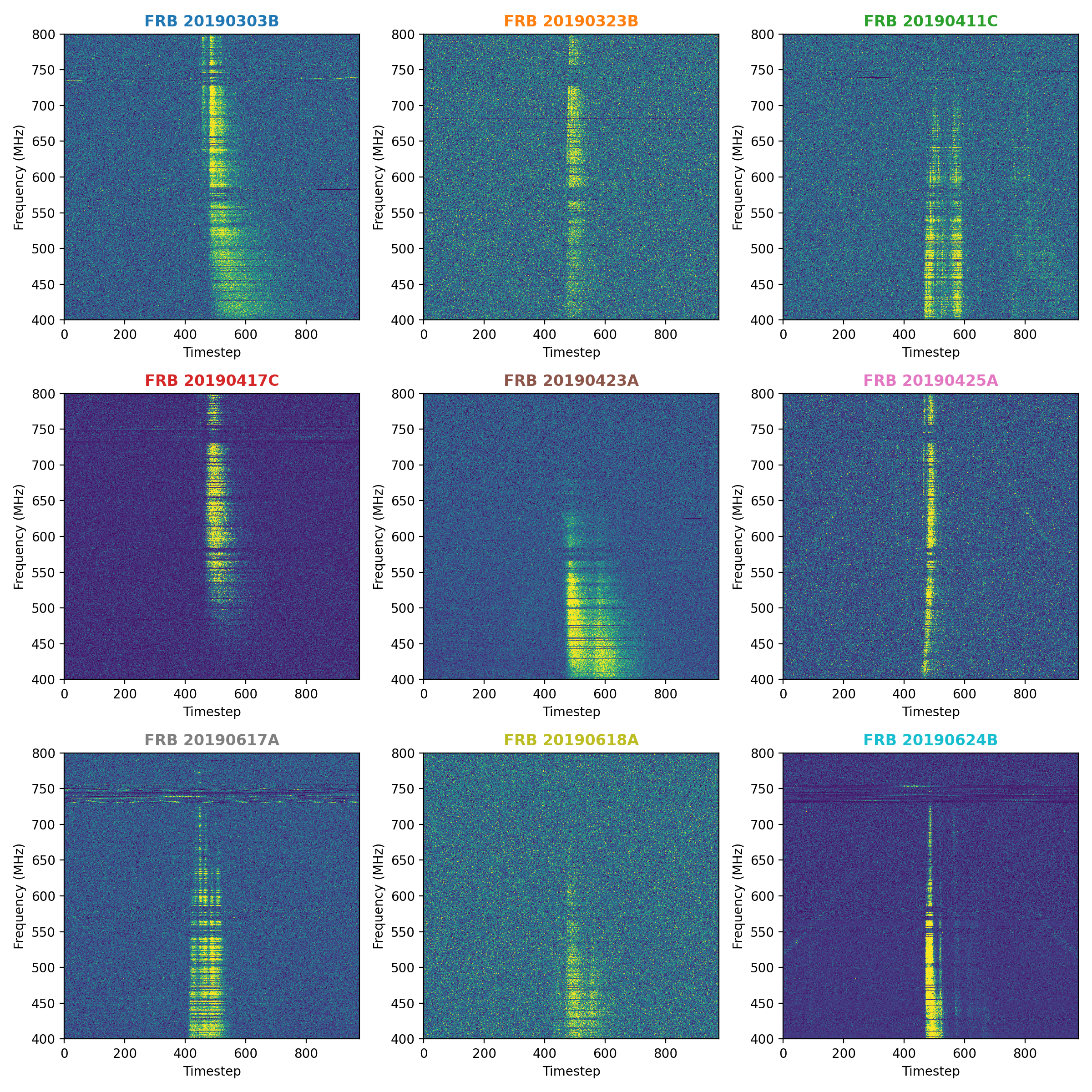}
    \caption{Dynamic spectra of the nine identified outliers from the CHIME dataset, showing unusual spectral characteristics. Some of these bursts exhibit complex morphology, including multiple components or pronounced scattering tails, while others, such as FRB 20190417C and FRB 20190423A, display channelization artifacts, visible as sweeping stripes across frequency channels. This highlights the importance of robust outlier detection methods for both astrophysical classification and instrumental artifact identification.}
    \label{fig:pca_analysis2}
\end{figure*}

Interestingly, many of these outliers correspond to bursts with more complex morphologies, including multiple components or significant scattering tails, setting them apart from the majority of bursts, which cluster in the low PC1 and PC2 regime and exhibit simpler, single-component, Gaussian-shaped structures. As shown in Figure~\ref{fig:pca_analysis2}, these outliers display diverse temporal and spectral structures, reinforcing the idea that PCA is not only effective in identifying statistical anomalies but also in revealing physically distinct burst morphologies within the dataset.

To further investigate these anomalies, we examined the dynamic spectra of the identified outliers (Figure~\ref{fig:pca_analysis2}). This revealed that some bursts, such as FRB 20190417C and FRB 20190423A, exhibited channelization artifacts—spurious features introduced by the instrument’s response and data processing pipeline. These artifacts, which appeared as sweeping stripes across the dynamic spectrum, are present to some degree in all bursts but become visually apparent primarily in the brightest ones. If not properly identified, such artifacts could mislead interpretations of the bursts.

The results demonstrate that PCA is effective in detecting outliers and highlighting anomalous features in large datasets, which might otherwise be missed through manual inspection. Importantly, PCA has identified bursts with complex astrophysical structures, such as multiple components or significant scattering, which differentiate them from the more common single-component Gaussian bursts. The inclusion of multiple outliers strengthens the conclusion that PCA serves as a useful tool for both exploratory data analysis and astrophysical burst classification, enabling systematic identification of unusual bursts and potential instrumental effects.

\subsection{IOB-CAE Reconstruction}

Building on the limitations observed in PCA reconstructions (Section~\ref{sec: results1}), we turn to the IOB-CAE to evaluate its ability to reconstruct both simulated and real CHIME bursts, including their complex morphologies and realistic noise characteristics. 

\begin{figure*}
    \centering
    \includegraphics[width=\textwidth]{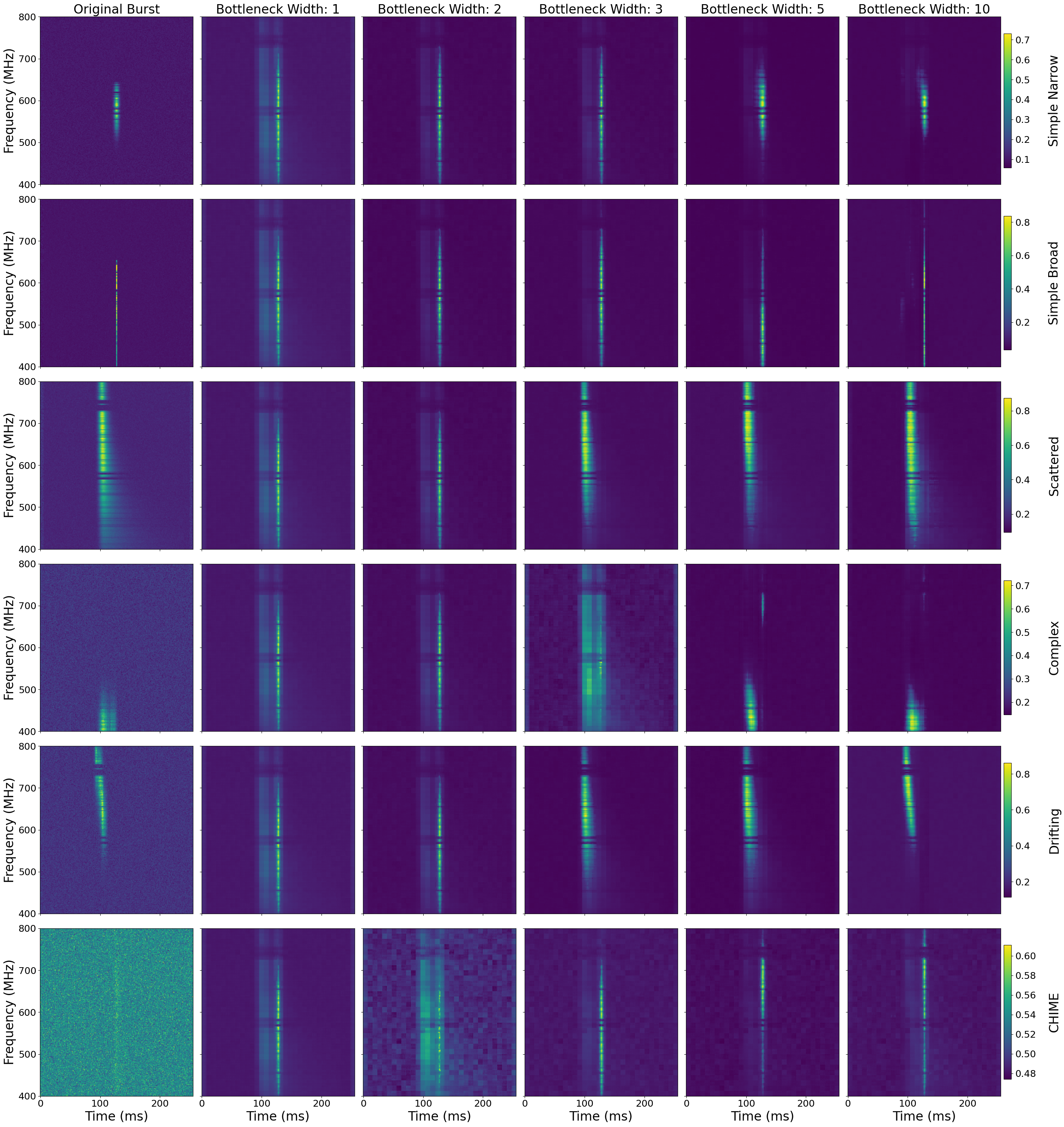}
    \caption{Reconstruction quality of IOB-CAE using real CHIME bursts. Each row represents a different burst type (Simple Broad, Simple Narrow, Scattered, Complex, Drifting and CHIME). The first column shows the original CHIME bursts, while each subsequent column shows reconstructions with an increasing number of latent variables (from one to ten).}
    \label{fig:reconstructions_real}
\end{figure*}

Figure~\ref{fig:reconstructions_real} illustrates the qualitative reconstruction progression of the IOB-CAE for various burst classes, with a particular focus on its ability to generalize to real CHIME bursts. Starting with a single latent variable, the reconstructions produce generalized and blurred versions of the bursts, similar to PCA with very few components. In this regime, the model lacks sufficient latent dimensions to encode detailed information, resulting in reconstructions that capture only the most basic global properties of the bursts, such as average intensity and coarse morphology. This explains why the second column appears similar across different burst classes; at this stage, the IOB-CAE focuses only on representing the dominant, low-resolution features common to all bursts, such as their general energy distribution.

As the number of latent variables increases, the reconstructions progressively improve, capturing key features such as frequency locations, burst shapes, and finer structures. By the third column (two latent variables), the reconstructions begin to exhibit subtle distinctions between burst classes, though some overlap remains due to the model’s focus on high-level structural features. Most of the essential details are effectively reconstructed by five latent variables, while reconstructions with ten latent variables closely resemble the original bursts across all types.

The reconstructions in Figure~\ref{fig:reconstructions_real} also demonstrate the IOB-CAE’s robustness in denoising bursts, even at realistic signal-to-noise (S/N) levels. Noise, which often dominates FRB observations, is effectively filtered out while preserving the core signal characteristics. This is evident in the bottom row of Figure~\ref{fig:reconstructions_real}, which shows reconstructions of CHIME bursts. Despite the limited number of real CHIME bursts relative to the simulated dataset, the IOB-CAE captures their underlying structure with notable accuracy. For example, narrow-band CHIME bursts are reconstructed with high fidelity, emphasizing the model’s ability to generalize across diverse data sources.

Notably, the IOB-CAE excels at handling complex morphologies, such as scattered and drifting bursts, where PCA tends to fail. This capability is especially important when working with real-world data, as it ensures that even intricate features of FRBs are preserved during reconstruction. Additionally, the IOB-CAE's ability to retain key features with relatively few latent variables highlights its efficiency in balancing compression and reconstruction quality.

These results underscore the superiority of the IOB-CAE in reconstructing FRB signals, particularly when applied to real-world data like CHIME bursts. As more real bursts are incorporated into the training set, we anticipate further improvements in the model’s ability to generalize.

\section{Conclusions and Outlook}
\label{sec:conclusions_outlook}

This study investigated the use of both traditional dimensionality reduction (PCA) and deep learning techniques (specifically, an IOB-augmented convolutional autoencoder) for analyzing FRB dynamic spectra. Through PCA, we conducted an initial exploration of the FRB data, examining separability among burst types and identifying outliers. Although PCA provided useful insights, its limitations in handling complex, non-linear structures became apparent. In contrast, the IOB-CAE model more effectively captured these intricate patterns, demonstrating improved signal fidelity and feature representation. Below is a summary of our findings and directions for future work.

\subsection{Conclusions}

\begin{itemize}
    \item \textbf{PCA as a Baseline for Initial Exploration}: PCA served as a valuable tool for initial data exploration, offering a straightforward, computationally efficient method to assess broad trends and separability in the dataset. It successfully captured simple structural features and enabled effective outlier detection, which is crucial for managing large, diverse datasets. However, the limited dimensionality and linear nature of PCA restricted its ability to reconstruct complex burst types accurately. PCA's utility could extend further with larger datasets, especially for identifying potential trends or anomalies at scale.
    
    \item \textbf{Enhanced Reconstruction with IOB-CAE}: The IOB-augmented CAE achieved high reconstruction quality across both simple and complex FRB types, significantly outperforming PCA in capturing the nuanced, non-linear structures of complex bursts. The IOB layer enabled efficient data compression by prioritizing essential information, allowing accurate reconstructions with a minimal number of latent variables. This result highlights the model’s capability to preserve critical signal features, even in complex burst morphologies. Importantly, the IOB-CAE generalized well to real CHIME bursts, demonstrating its ability to capture key shared features despite the smaller sample size of real data.
    
    \item \textbf{Effective Denoising and Signal Fidelity}: The IOB-CAE demonstrated robust denoising capabilities, even at realistic signal-to-noise ratios, producing clearer and more accurate representations of the original bursts as latent variables increased. This capability is particularly notable given the relatively small dataset of $\sim$5,000 simulated bursts used for training. While this dataset size highlights the feasibility of applying deep learning methods to current FRB catalogs, it also underscores the potential for improved performance with larger, more diverse datasets. Incorporating additional real and simulated data could enhance the model’s ability to adapt to complex variations and subtle features.
    
    \item \textbf{Latent Space Insights and Generalization}: Initial separability observed in the IOB-CAE’s latent space suggests that the model can capture intrinsic differences between FRB types. However, the observed continuum of burst morphologies within the latent space suggests that FRBs may not naturally form discrete clusters, but instead exhibit overlapping characteristics. This continuum aligns with the current understanding of FRB diversity and morphology. Furthermore, future work could explore the use of latent space structures to study differences between repeaters and non-repeaters, providing additional insights into FRB populations.
\end{itemize}

In summary, the combination of PCA for preliminary exploration and IOB-CAE for detailed reconstruction provides a comprehensive framework for FRB signal analysis, balancing efficiency and detail preservation. PCA offered a useful starting point, particularly for outlier detection and simple structure identification, while the IOB-CAE excelled in capturing the complex, non-linear features necessary for a nuanced understanding of FRB morphologies.

\subsection{Outlook}

\begin{itemize}
    \item \textbf{Expansion with Larger and More Diverse Datasets}: As FRB datasets grow in volume and diversity, notably with new CHIME/FRB catalog releases, the robustness of both PCA and IOB-CAE should be reassessed on these larger datasets. Increasing the number of training examples could enhance both methods' ability to generalize, particularly for the IOB-CAE, which may improve its latent space organization and class separability. Furthermore, the growing availability of data offers an opportunity to explore differences between FRB repeaters and non-repeaters using latent space representations.
    
    \item \textbf{Incorporating Advanced FRB Models}: Expanding the \texttt{FRBakery} simulation tool to encompass more complex morphologies, such as intricate frequency-time variations and microstructure, will enable more realistic training for the autoencoder. These enhancements will introduce significantly higher dimensionality, as capturing fine temporal structures will require considerably more data points. For instance, if a burst lasts for 10~ms and exhibits structure on 1~$\mu$s timescales, at least 10,000 time bins would be necessary to fully resolve the burst. Handling this increased dimensionality will be essential for the IOB-CAE to effectively adapt to the diversity expected in future FRB observations.

    \item \textbf{Alternative Loss Functions for Improved Reconstructions}: Investigating alternative loss functions, such as the Structural Similarity Index (SSIM) \citep{SSIM}, may further improve the IOB-CAE’s ability to preserve structural details, potentially leading to better reconstructions and denoising performance. Such enhancements could also improve model interpretability by maintaining essential features of the bursts.
    
    \item \textbf{Latent Space Interpretability and Regularization}: Interpreting the latent space is essential for understanding the features captured by deep learning models. Future work could incorporate techniques such as those presented in \cite{lucie2024deep}. This approach involves an interpretable variational encoder \citep{Kingma2019AnIT} that returns the independent factors of variation within the latent space, providing insights into the model's decision-making process. By identifying and understanding these factors, we could gain deeper insights into what the latent components physically represent.
    
    \item \textbf{Leveraging Simulated and Real Data Synergy}: While the current study demonstrates the feasibility of combining real and simulated data, further exploration into how the interplay between these datasets affects latent space organization and reconstruction quality will be critical. Specifically, identifying whether simulated bursts bias the latent space structure or whether real bursts anchor the latent space to observed FRB morphologies will provide valuable insights for designing future training datasets.
\end{itemize}

\section*{Acknowledgements}

DK and JWTH acknowledge funding for this work from the European Research Council (ERC) under the European Union’s Horizon 2020 research and innovation programme (ERC Advanced Grant `EuroFlash'; Grant agreement No. 101098079). The AstroFlash research group at McGill University, University of Amsterdam, ASTRON, and JIVE is additionally supported by: a Canada Excellence Research Chair in Transient Astrophysics (CERC-2022-00009) and an NWO-Vici grant (`AstroFlash'; VI.C.192.045). DH acknowledges support from NWO's Women In Science Excel (WISE) programme.
GC acknowledges support from the European Union's Horizon Europe research and innovation programme under the Marie Skłodowska-Curie Postdoctoral Fellowship Programme, SMASH co-funded under the grant agreement No. 101081355. The authors wish to acknowledge the Institute for Fundamental Physics of the Universe (IFPU), in Trieste, for hosting and supporting this work through their Team Research program. The authors thank Matt Ho for insightful discussions regarding the IOB. The authors thank Ziggy Pleunis for the helpful comments on the manuscript.

\section*{Data Availability}

All the complex voltage data used are sourced from the \href{https://www.chime-frb.ca/}{CHIME/FRB catalogue}. The simulated data used in this study are available on \href{http://doi.org/10.5281/zenodo.14337997}{Zenodo}. The code used for the analysis and simulations are available on the \href{https://github.com/SRON-API-DataMagic/Rep_Learn_FRB/tree/main/FRBakery}{FRBakery github page}.



\bibliographystyle{mnras}
\bibliography{mnras_template} 



\appendix


\bsp	
\label{lastpage}
\end{document}